\documentclass[a4paper,fleqn]{cas-dc}

\newtheorem{definition}{Definition}%

\usepackage{natbib}
\setcitestyle{numbers,square}

\usepackage{threeparttable}
\usepackage{multirow}
\usepackage{makecell}
\usepackage{float}
\usepackage{hyperref}
\usepackage{amsmath, amssymb}
\usepackage{makecell}
\usepackage{multirow}
\usepackage{algorithm,algpseudocode,float}
\usepackage{lipsum}
\usepackage{comment}
\usepackage{diagbox} 
\usepackage{xcolor}
\usepackage{upgreek}
\newcommand{\tabincell}[2]{\begin{tabular}{@{}#1@{}}#2\end{tabular}}

\makeatletter
\newenvironment{breakablealgorithm}
  {
   \begin{center}
     \refstepcounter{algorithm}
     \hrule height.8pt depth0pt \kern2pt
     \renewcommand{\caption}[2][\relax]{
       {\raggedright\textbf{\ALG@name~\thealgorithm} ##2\par}%
       \ifx\relax##1\relax 
         \addcontentsline{loa}{algorithm}{\protect\numberline{\thealgorithm}##2}%
       \else 
         \addcontentsline{loa}{algorithm}{\protect\numberline{\thealgorithm}##1}%
       \fi
       \kern2pt\hrule\kern2pt
     }
  }{
     \kern2pt\hrule\relax
   \end{center}
  }

\begin{document}
\let\WriteBookmarks\relax
\def\floatpagepagefraction{1}
\def\textpagefraction{.001}

\title {\LARGE{BPCE: A Prototype for Co-Evolution between Business Process Variants through Configurable Process Model}}

\author[1]{Linyue Liu}
\author[1]{Xi Guo}
\author[2]{Chun Ouyang}
\author[3]{Patrick C. K. Hung}
\author[1]{Hong-Yu Zhang}
\author[4]{Keqing He}
\author[5]{Chen Mo}
\author[1]{Zaiwen Feng*}[
orcid=0000-0003-1618-3553,
]

\address[1]{College of Informatics, Huazhong Agricultrual University, Wuhan 430070, China}
\address[2]{School of Information Systems, Queenland University of Technology, Brisbane 4000, Australia}
\address[3]{Faculty of Business and Information Technology, Ontario Tech University, Oshawa L1G0C5, Canada
}
\address[4]{School of Computer, Wuhan University, Wuhan 430072, China}
\address[5]{Wenbo Investment Management Co.Ltd., Shanghai 200120, China}

\cortext[cor1]{Corresponding author.}

\begin{abstract}
With the continuous development of business process management technology, the increasing business process models are usually owned by large enterprises. In large enterprises, different stakeholders may modify the same business process model. In order to better manage the changeability of processes, they adopt configurable business process models to manage process variants. However, the process variants will vary with the change in enterprise business demands. Therefore, it is necessary to explore the co-evolution of the process variants so as to effectively manage the business process family. To this end, a novel framework for co-evolution between business process variants through a configurable process model is proposed in this work. First, the mapping relationship between process variants and configurable models is standardized in this study. A series of change operations and change propagation operations between process variants and configurable models are further defined for achieving propagation. Then, an overall algorithm is proposed for achieving co-evolution of process variants. Next, a prototype is developed for managing change synchronization between process variants and configurable process models. Finally, the effectiveness and efficiency of our proposed process change propagation method are verified based on experiments on two business process datasets. The experimental results show that our approach implements the co-evolution of process variants with high accuracy and efficiency.
\end{abstract}

\maketitle

\section{Introduction}\label{sec1}

In the past two decades, business process management technology has developed rapidly and has been widely used in workflow systems  \cite{r18} and service computing fields  \cite{r15,r16}. The co-existence of multiple variants of the same business process is a widespread phenomenon in contemporary organizations. As a concrete example, The Netherlands has around 430 municipalities, which in principle execute the same or very similar set of processes \cite{rosa2017business}. In the context of company mergers and restructurings, multiple variants of the same process often occur in a Cloud, usually originating from different companies or units, which need to \emph{co-evolve} \cite{r1}. Moreover, a large number of subsidiaries of a large-scale organization may need to evolve the business processes at the same time according to the orders of their company, such as China Mobile Communications Corporation (CMCC) \cite{r48}. These multiple \emph{process variant}s, which are collectively known as a \emph{process family}, probably run in different BPM engines in parallel, making it hard to co-evolve them in a centralized way \cite{r25}. Therefore, the methods and tools which can ensure each process variant in a process family co-evolves in terms of the demand of co-evolution to help reduce costs and avoid inconsistencies.

The objective of this study is to achieve the synchronized change propagation between the process variants to make the whole process families co-evolve. Several previous studies  \cite{r2,r6,r7} have proposed the co-evolution methods of the business process model family. For example, an aspect-oriented technology (AOP)  \cite{r34} is applied for managing the co-evolution of process families  \cite{r2}. Based on the revision history of a process repository, an approach for dependency-based impact analysis of the business process repository was proposed  \cite{r6}. Another approach based on behavioural profiles of corresponding activities has been proposed to determine a change region for change propagation between two semantically overlapping process models  \cite{r7}. However, the problem with the state-of-the-art methods is that the cost of co-evolving is still too expensive. Semantic annotations on AOP-based \emph{plug-in}s \cite{r2} or \emph{activities} of the business process model \cite{r7} are compulsory preliminary work with the support of inferences by domain ontologies. Alternatively, change primitives are defined in Petri Net representation for the process model \cite{r6,r25}, which is also time-consuming, especially when the number of variants is considerable. This issue of the state-of-the-art affects significantly the efficiencies and effectiveness of Cloud process family co-evolution.

To address the problem, we advocate the use of a \emph{configurable process model} as a mediator for co-evolving process families in this work. A configurable process model includes a family of process variants in an integrated manner, which allows analysts to understand the commonalities and differences of these process models and possible reasons for the differences  \cite{r4}. The major contributions of this work are threefold:

\begin{itemize}
    \item We define \emph{change primitives} to describe change operations of process variants and configurable process models. Subsequently, \emph{change propagation operations} are defined from process variants to configurable process model and from a configurable process model to process variants.
    \item An overall change propagation algorithm is proposed based on change propagation operations between process variants and a configurable process model. The experiments are conducted on 155 commercial BPEL instance models and 604 R/3 business process models of SAP  \cite{r9}. The results show that our method can achieve change propagation with high accuracy and extremely low time consumption.
    \item We have implemented all the change operations and the change propagation algorithms and then integrated them into APROMORE\footnote{https://github.com/apromore}(An Advanced Process Model Repository, APROMORE  \cite{r5}), an international large-scale open-source advanced business process warehouse.
\end{itemize}

The paper is structured as follows: Section 2 shows the related work of the research; Section 3 provides some preliminary definitions; Section 4 explains the algorithm and presents the workflow; Section 5 describes the architecture of the prototype; Section 6 evaluates the feasibility and efficiency of our proposed method; Finally, Section 7 is the conclusion.

\section{Related Work}\label{sec6}

The following subsections discuss related work directly addressing change propagation between process models as well as the approaches which are potentially useful in the co-evolution of process models.

\subsection{Co-Evolution of Business Processes}

The co-evolution of business process families has received extensive attention in the field of business process management in recent years. 

Grossmann \emph{et al.} (2013) propose a change propagation framework working between the Common Reference Model and the Process View and a consistency detection method between different views \cite{r25}\cite{r3}.

Feng \emph{et al.} (2017) propose an \emph{aspect-oriented-programming} based business process family co-evolution method, which encapsulates the co-evolution information into a plug-in and inserts the plug-in into the business process family through aspect-oriented technology to realize the co-evolution of process family \cite{r2}.

Weidlich \emph{et al.} (2012) propose a process model family change propagation method based on the behaviour profile has been proposed in another study, in which when a process in the processing warehouse is changed, the changes of the process behaviour profile are extracted, and the changing area of the related process is determined to realize the co-evolution of the processes in the entire business warehouse \cite{r7}.

Weber \emph{et al.} (2011) propose a method to propagate changes of a configurable process model to already configured process variants based on "smell", which can remove unused paths from a process model and generalize frequently occurring instance changes by pulling them up to the process type level when some variants are changed \cite{weber2011}.

Song and Jacobsen (2016) propose a framework for business process change management  \cite{r11}, and under the same framework, our method design business process family co-evolution. 

An approach for analyzing the revision history of a process repository has been proposed \cite{r6}. In this approach, the change impact is computed based on the view that business processes which co-vary in the past are likely to change together in the future. The preconditions and effects of activities in business process models are firstly semantically annotated, and then the impact of collaborative changes between processes and sub-processes are analyzed by calculating cumulative effects and a dependency-based impact analysis technology. Unlike our method, this method requires a great amount of preparatory work for semantic annotation of process models and revision history generation of a process repository. These two requirements may not be satisfied even in a state-of-the-art process model repository.

\subsection{Business Process Variability}

Poizat \emph{et al.} (2016) propose a range of business process evolution methods to change the variants, but these methods only involve the evolution of one single business process and are not suitable for the entire family of business processes, and only relate to model variability \cite{r47}.

Arellane and Lau (2019) propose a method to change workflows at runtime for different contexts in the field of IoT \cite{r40}. This method combines variability with behaviour and offers an infinite number of workflow variants but is based on DX-MAN \cite{r41} and not suitable for BPMN.

Dynamic business processes (DBP) have been studied over the past years. Cognini \emph{et al.} (2018) present a literature review on BP's dynamic and define dynamic as the ability to manage the coordination between challenges in organizational aspects and technical environments \cite{r43}. DBP supports structural and functional changes according to its context and rules \cite{r42}, which is similar to the configurable model. A web application named business process family manager(BPFM) is proposed to manage BP family. BPFM includes a meta-model to define BP relationship and version, can generate process variants from basic processes according to user requirements \cite{r36}. 

Similarly, Calegari \emph{et al.} (2019) propose a method based on common variability language (CVL), these method uses the common and variable parts to automatically produce a process variant from the BP family by fragment substitution, and improves the coverage of the process views \cite{r39,r38}. 

To solve the configuration problem of BPFM, {\v{S}}endelj and Ognjanovi{\'c} (2018) propose a method with analytic hierarchy process to address different kinds of preferences and derive process variants with behavioural correctness from a business process family \cite{r44}. Furthermore, a method based on process performance indicators (PPIs) is proposed to generate variants between PPIs variability and other processes perspectives \cite{r46}. This method is suitable for restrictive and extensible variability, formalizes how variants are derived for each variability, and defines the conditions that the variability model must satisfy to obtain syntactically correct variants. The above methods are similar to our method and related to business process configuration which includes change propagation, generating variants and ensuring correctness.

\subsection{Business Process Versioning}

Thomas (2008) proposed a technique to support the co-existence of different versions of the same configurable process model in the context of long-running processes by version-graph models, which can realize the management of different versions and get the changes between them \cite{thomas2008}.

Gerth \emph{et al.} (2013) propose a method for detecting and resolving conflicts between different versions of the models  \cite{r22}, and their method distinguishes semantic conflicts from grammatical conflicts and uses terminology to avoid misjudgment of most grammatical conflicts, but its conflict resolution strategy is not suitable for our method. 

Brosch \emph{et al.} (2010) propose a modeling tool to identify model reconstruction and apply this identification method to conflict detection between different versions of the model, and they further propose a conflict resolution method based on predefined patterns, which contributes to model conflict detection  \cite{r23}.

Song \emph{et al.} (2021) propose a search algorithm that combines heuristics and $A^*$  \cite{r20} to determine the sequence of minimal change operations for the process model \cite{r8}. Li \emph{et al.} (2008) propose a matrix-based method to detect the minimum change sequence \cite{r17}, and Gerth \emph{et al.} also propose a method to detect model discrepancies without change logs \cite{r21}. These studies are conducive to propagating a complex set of change operations to configurable business process models and other variants.

The format of SAP reference model dataset is \emph{Event-driven Process Chains}(EPC) format. Dreiling \emph{et al.} (2005) propose an algorithm to individualize the C-EPC graph into a regular EPC graph, and if C-EPC graph is not syntactically correct, this algorithm deletes all the nodes that are not on the path between the start and end nodes and then connects the remaining nodes  \cite{r31}, which is partially similar to the cleaning operations removing redundant paths in our method. 

Dreiling \emph{et al.} (2006) propose that a configurable connector can be restricted to any node subset on its incoming or outgoing edges  \cite{r32}. For example, an AND split connector can be configured as a regular AND connector but with a limited set of output edges. 

Marcello \emph{et al.} (2011) propose the C-iEPC (Configurable integrated EPC) language  \cite{r33}, which supports specifying configurable nodes in organizational resources and object classes in business processes. The above processing method of EPC map contributes to converting the EPC format.

\subsection{Fragment customization}
Approaches in this group are based on the application of \emph{change operations} to restrict or extend the configurable process model \cite{rosa2017business}. Two atomic change operations can be used to customize the control flow: \emph{delete}, remove a fragment from the model, and \emph{insert}, add a fragment into the model.

Hallerbach \emph{\emph{et al.}} propose a process configuration method by applying change operations to a \emph{reference model} marked with adjustment points \cite{hallerbach2010capturing}. In this method, the reference model is a standard process, the most frequently used process variant, a generic model, but without any variation points. Provop supports four operations (i.e. DELETE, MODIFY, MOVE and INSERT), to delete, change and relocate a fragment delimited by two adjustment points, or insert a new fragment to another part of the model delimited by two adjustment points in the reference model.

Kumar \emph{\emph{et al.}} propose a method for process family variability management by processing a series of \emph{business rules} associated with a \emph{process template} \cite{kumar2012design}. The rules are sequences of change operations used to configure the template by restricting or extending its behaviour. However, configuration rules are not graphically represented in any process model perspective.

\section{Preliminaries}
Our process family co-evolution technique includes the following four basic preliminaries, namely, configurable process model notation, business process variant merging technique, change operations on process graphs, and detection of change operations.

\subsection{Configurable business processes}

In the context of company merging and restructuring, multiple business process variants which are usually originated from different companies and their sections need to co-evolve and eventually converge into a single process in order to eliminate redundancies and create synergies  \cite{r1}. To this end, business analysts often compare the existing process variants to identify their commonalities and differences, thus generating a unified business process model to promote process integration. The unified process model is called the Configurable Process Model \cite{r4}.

There exist many notations to represent business processes such as event-driven process chains (EPC), UML activity diagrams (UML ADs) and the business process modeling notation (BPMN). In this study, we adopt a directed graph with labeled nodes to define a business process model, as described in a previous study \cite{r1}.

\begin{definition}[\textbf{Business Process Graph $G$ \cite{r1}}] 
A \emph{business process graph} $G$ is a set of pairs of process model nodes, i.e.,\ each pair denoting a directed edge. A node $n$ of $G$ is a tuple $(id_G(n)$, $\lambda_G(n)$, $\tau_G(n))$ consisting of a unique identifier $id_G(n)$ within $G$, a label $\lambda_G(n)$, and a type $\tau_G(n)$. 
\end{definition}

In general, there are three types of nodes: \emph{function} nodes representing tasks that can be performed in an organization, \emph{event} nodes representing pre-conditions that must be satisfied before the node function is performed, or post-conditions that must be satisfied after node function has been performed; and \emph{connector} nodes determining the execution flow of the process. In the formula $\tau_G\in$ \{`$f$', `$e$', `$c$'\}, the letters ‘f’, ‘e’, and ‘c’ represent the (\emph{f})unction, (\emph{e})vent and (\emph{c})onnector node type, respectively.

Additional connector nodes are often required when business process variants are merged into a configurable business process model. Thus, these additional connector nodes only exist in the merged configurable business process model rather than in any of the process variants. We name these additional nodes \emph{auxiliary connector nodes}.

\begin{definition}[\textbf{Configurable Business Process Graph $CG$\cite{r1}}] 
Let $\mathcal{I}$ be a set of identifiers of business process graphs, $\mathcal{N}$ a set of node identifiers in business process graphs, and $\mathcal{L}$ the set of all possible node labels. A \emph{configurable business process graph} $CG$ is a tuple $(G^\ast$, $\alpha_{G^\ast}$, $\beta_{G^\ast}$, $\eta_{G^\ast})$ where  
	\begin{itemize}
		\item $G^\ast$ is a business process graph,
		\item $\alpha_{G^\ast}: G^\ast\rightarrow 2^\mathcal{I}$$\setminus$$\emptyset$ is a function that maps each edge in $G^\ast$ to a set of process graph identifiers, 
		\item $\beta_{G^\ast}: N_{G^\ast}\rightarrow 2^{\mathcal{I}\times\mathcal{N}}$ is a function that maps a node $n'\in N_{G^\ast}$ to a set of pairs $(\mathit{pid},\mathit{nid})$ where $\mathit{pid}$ is a process graph identifier and $nid$ is the identifier of node $n'$ in process graph $G_\mathit{pid}$.\footnote{Such information can be recorded during process merge.} A node $n'\in N_{G^\ast}$ is an auxiliary connector node if and only if $\beta_{G^\ast}(n')=\emptyset$,
		\item $\eta_{G^\ast}: N_{G^\ast}\rightarrow\{\mathit{true},\mathit{false}\}$ is a boolean function indicating whether a node is configurable or not.
	\end{itemize}
\end{definition}

\begin{figure*}
    \centering
    \includegraphics[width=150mm]{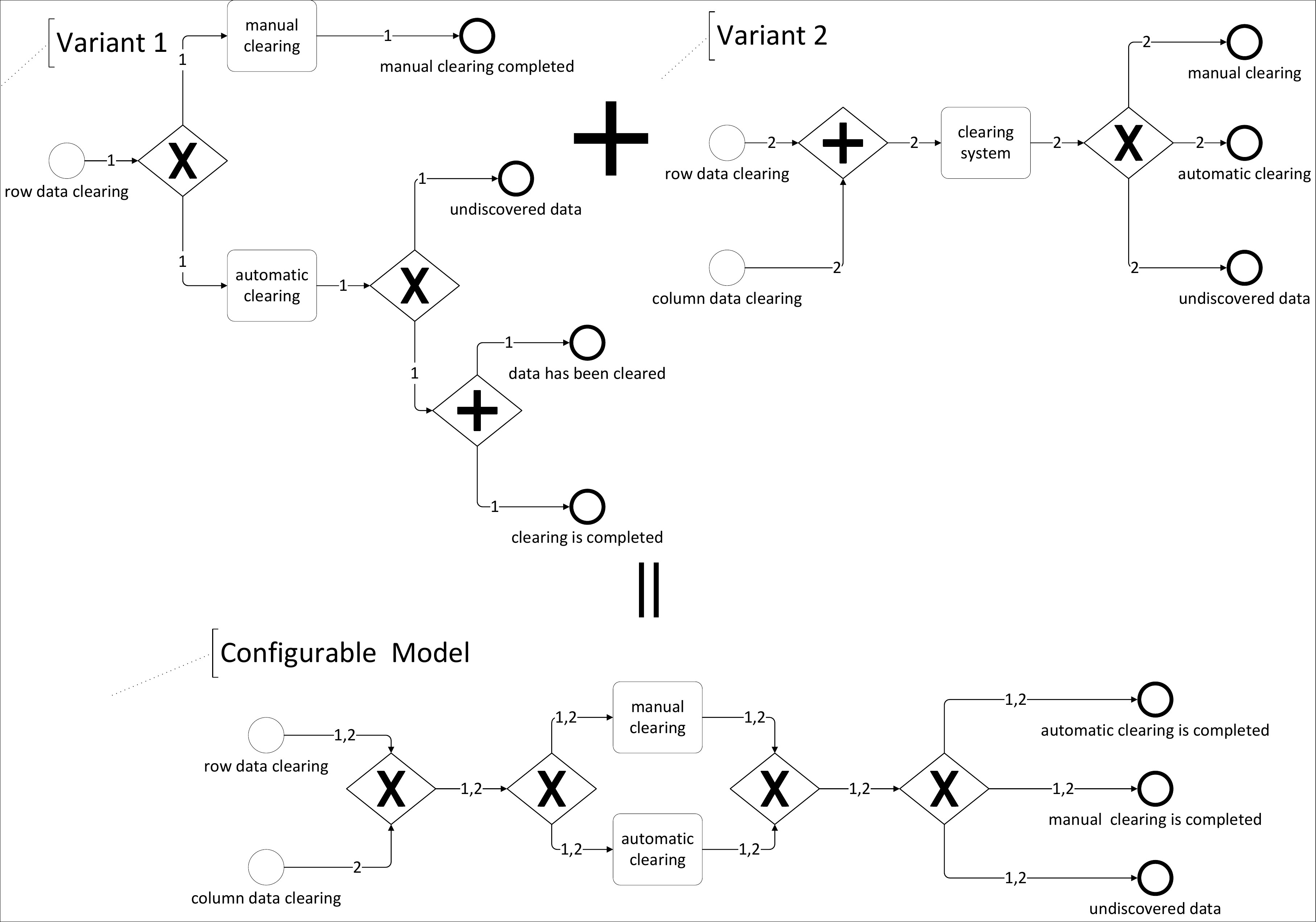}
    \caption{ SAP Reference process model for EXCEL data processing \cite{r9} }
    \label{SAP reference model}
\end{figure*}

Here, we selected "Excel Data Processing" from SAP R/3 reference model \cite{r9} as a simplified version of the reference business process model. The reference process model has two \emph{process variants}, as shown at the top of Fig. \ref{SAP reference model}. Process variant 1 describes the processing method of row data which is divided into the manual clearing and automatic clearing. Manual clearing can be ended directly, while automatic clearing is further divided into two cases, namely, undiscovered data and discovered data. Once data is found, the clearing process is completed, namely, the data has been cleared. In process variant 2, row data and column data can be processed simultaneously by a clearing system in which three final events occur, including manual clearing, automatic clearing, and undiscovered data. The process variants 1 and 2 are shown in the upper left/right corner of Fig. \ref{SAP reference model}, respectively.

\subsection{Merging business processes}

Before propagating the changes from one process variant to the other, we need to merge these process variants into a configurable process model. Manually merging process variants is a tedious and error-prone task. To solve this problem, a typical method has been proposed  \cite{r1}, and this method starts with creating an initial version of the merged graph $CG$ of two process variants $G_1$ and $G_2$, by computing of the union set of the edges of $G_1$ and $G_2$. Next, the mappings between $G_1$ and $G_2$ are partitioned into maximum common regions (\emph{mcrs}). At last, the \emph{mcrs} are connected with the remaining nodes that are not matched by $G_1$ and $G_2$ to generate the final merged graph $CG$.

Further, we summarize the approach of business process consolidation  \cite{r1}. First, we calculate the node similarity between variants by computing the \emph{label} similarity of the events and functions, and the \emph{context} similarity of the connectors (based on their maximum common parent and child nodes). Not every node and edge in the two variants can be mapped, and those nodes (edges) that cannot be mapped will be directly inserted or deleted when merging, thus, they are called "inserted or deleted nodes (edges)." Afterward, the "similarity" and "inserted or deleted nodes and edges" are weighed and then added to obtain the final matching score. The mappings with the highest matching score were taken as the final mappings between process variants. After mapping establishment, we merge the nodes and edges in the mapping, which is called "Merging Maximum Common Regions", and then we directly insert the nodes and edges that have not been mapped into a merged model with some unmapped nodes and edges discarded. Finally, the illegal parts in the merged model are removed through a cleaning operation. The details of the merging algorithm are available in the literature \cite{r1}.

\subsection{Change operations}\label{subsec3.1}

Before co-evolving business process variants through their configurable process model, we need to establish a set of \emph{change operation}s on process variants and configurable process model, respectively, to perform a set of update operations on a graph. Referring to change operations defined in the \emph{on-line graph maintenance problem}  \cite{r19}, where the updates are insertions of vertices and edges, the change operations of process variants and configurable process models are defined as follows.

\subsubsection{Change operations of variants}\label{subsubsec3.1.1}

 We first define the 2 primitive operations (i.e., (1) and (2)) on a business process graph $G$. \emph{Primitive} operations refer to those basic operations which are independent of any other operation for the change of $G$. Then, based on above-defined primitive change operations, 4 change operations (i.e. (3)-(6)) are defined. Change operations represent those operations based on the primitive operations for the change of $G$. The 6 change operations are as follows.

\begin{enumerate}
\renewcommand{\labelenumi}{(\theenumi)}
\item "Insert Edge": inserting a new edge between two nodes of the variant.
\item "Delete Edge": deleting the edge between the two nodes of the variant and releasing the parent-child relationship between the nodes.
\item "Insert Node": inserting a new node onto the edge of the two nodes by adding the new path first and then deleting the old edge.
\item "Add Node": adding a single-node path between two connectors.
\item "Append/Prepend Node": adding a node in front of / behind a node. Since these two operations are similar, we combine them into one operation.
\item "Modify Node Annotation": modifying the non-null annotation of a node.
\end{enumerate}

Detailed information on the definitions of each primitive or change operation is presented in Appendix B.

\subsubsection{Change operations of configurable process models}\label{subsubsec3.3}

Next, we define the 8 change operations (i.e., (1)-(8)) on a configurable business process graph $CG$. The detailed definitions of each primitive or change operation on $CG$ are presented in Appendix C.

\begin{enumerate}
\renewcommand{\labelenumi}{(\theenumi)}
\item "Insert Edge": inserting a new edge between two nodes of the configurable process model.
\item "Delete Edge": deleting the edge between the two nodes of the configurable process models and releasing the parent-child relationship between the nodes.
\item "Insert Node": inserting a new node on the edge of the two nodes.
\item "Add Node": adding a single-node path between two connectors.
\item "Append/Prepend Node": adding a node in front of /behind a node. These two operations are similar, and thus we combine them into one.
\item "Modify Node Annotation": modifying the non-null annotation of a node.
\item "Insert Edge Annotation": inserting the new annotation onto an edge.
\item "Delete Edge Annotation": deleting the edge annotation of a certain edge. If the annotation is empty after the deletion, this edge will be deleted through the cleaning operations.
\end{enumerate}

The change operations of $G$ and $CG$ are the basis for change propagation operations between $G$ and $CG$.

\subsection{Detection of Change Operation Sequence}\label{subsubsec2.4}

Before change propagates between $G$ and $CG$, the sequences of change operations on $G$ or $CG$ must be detected and identified. To this end, when the user completes the modification on $G$ or $CG$ and saves it, the old and new version of process models are compared to obtain the sequence of change operations made by the user. In this article, we abstract from any specific notation and represent process change sequence as actions on a labeled graph as per the following definition.

\begin{definition}[\textbf{Process Change Sequence, updated from \cite{r17}}] Let \textsl{P} denote the set of possible process models and \textsl{C} the set of possible process changes. Let $S,S' \in \textsl{P}$ be two process models, let $\Delta \in \textsl{C}$ be a process change, and let $\sigma=<\Delta_1,\Delta_2,...,\Delta_n> \in \textsl{C}^*$ be a sequence of process changes performed on inital model $S$. Then:
\begin{enumerate}

	\item $S [\Delta\rangle S'$ iff $\Delta$ is applicable to $S$ and $S'$ is the process model resulting from the application of $\Delta$ to $S$.
	
	\item $S [\sigma\rangle S'$ iff $\exists S_1,S_2,...S_{n+1} \in \textsl{P}$ with $S=S_1$, $S'=S_{n+1}$, and $S_i [\Delta> S_{i+1}$ for $i \in {1,...n}$.

\end{enumerate}
\end{definition}

There are multiple methods to compare and obtain changes of different versions of the process model. By employing the method proposed by Li \emph{et al.} \cite{r17}, we obtained the minimal change sequences for changing the original model into a new version based on the change matrix. The conversion from user modification on the process model into the minimal sequence of change primitives consists of the following 3 steps:

\begin{itemize}
\item $N_{Delete} = N \backslash N' ; E_{Delete} = E \backslash E'$: All the edges that exist in the old model but not in the new model are deleted to obtain the "Delete Edge" primitive, and orphan nodes are removed through the cleaning operations.
\item $N_{Change} = N \cap N'$: The nodes in the old model are moved to their positions in the new model, and then their related edges are added or deleted. The modifications of the annotations of these nodes are examined. This step includes the primitives "Delete Edge," "Insert Edge," and "Modify Node Annotation." The algorithm reported by Li \emph{et al.} \cite{r17} uses the change matrix to reveal the movement conflicts between nodes in the model, and uses logical expressions to obtain the optimal solution for resolving the movement conflicts. Finally, the simplest method of moving nodes is obtained.
\item $N_{Add} = N' \backslash N ; E_{Add} = E' \backslash E$: The old model is added with the nodes and related edges existing in the new model. This step includes all the node and edge operation primitives. The old model will be completely turned into the new model at this step. Similarly, the conflict matrix is used to sort the added new nodes. 
\end{itemize}

After the user completes the modification, we compare the old and new models to obtain all the edges to be added or deleted and the nodes to be added or deleted, and then we use the above algorithm to convert these edges and nodes into the corresponding optimal change primitives. Subsequently, we invoke the propagation operation corresponding to the change primitive to complete the model change propagation.

\section{Approach}\label{sec3}

The co-evolution is defined over pairs of configurable process graphs (variants). In order to make two or more (non-configurable) process graphs co-evolve, we first need to convert each business process graph into a configurable process graph. This is trivially achieved by annotating every edge of a process graph with the identifier of the process graph, and every node in the process graph with a pair indicating the process graph identifier and the label for that node. We then present the basic change propagation operations. Next, we show how to propagate changes from a process graph (variant) to configurable process graphs and then, reversely, from configurable process graphs to process graphs. Finally, we discuss a set of cleaning rules to remove illegal parts in the merged process graph. The notation used in the algorithms of this article is summarized in Appendix A.

\subsection{Overall Algorithm}\label{subsec3.1}

Before co-evolving business process families through a configurable process graph, we need to establish which nodes in the process graph (variant) \emph{match} which nodes in the configurable process graph. Here, a \emph{mapping} is a function from the nodes in the process variant to those in the configurable process graph. We define the mapping of the nodes and edges between process variants and configurable process models in the following.

\begin{definition}[\textbf{Mapping between the nodes in $G$ and $CG$}]

Given a business process graph $G$, $\mathit{pid}_G\in\mathcal{I}$ is the process identifier of $G$. 
For each node $n\in N_G$, $\lambda_G(n)$ is the label of $n$. Given a configurable business process graph $CG$ that is obtained as a result of merging a set of business process variants $\mathcal{G}$ where $G\in\mathcal{G}$. 
Let $CG$ = $(G^\ast$, $\alpha_{G^\ast}$, $\beta_{G^\ast}$, $\eta_{G^\ast})$, $N_{G^\ast}$ is the set of nodes in~$CG$. 
The following two functions define mappings between nodes in a business process graph $G$ and a corresponding configuration business process graph $CG$. 
\begin{itemize}
	\item $\textsf{M}^\textsf{N}_\mathit{G,CG}: N_G \rightarrow N_{G^\ast}$ is an injective function that maps every node in $G$ to its corresponding node in $CG$, where $\textsf{M}^\textsf{N}_\mathit{G,CG} = \{(n,n')\in N_G\times N_{G^\ast} \ \vert \ \forall n\in$ $N_G$ $\exists n'\in N_{G^\ast}$ such that $\gamma_{G^\ast}(n')=(\mathit{pid}_G,\lambda_G(n))\}$.
  
	\item $\textsf{M}^\textsf{N*}_\mathit{CG,G}: N_{G^\ast} \nrightarrow N_{G}$ is an in partial function that maps certain node in $CG$ to its corresponding node in $G$, where $\textsf{M}^\textsf{N}_\mathit{G,CG} = \{(n,n')\in N_G\times N_{G^\ast} \ \vert \ \forall n\in$ $N_G$ $\exists n'\in N_{G^\ast}$ such that $\gamma_{G^\ast}(n')=(\mathit{pid}_G,\lambda_G(n))\}$.
\end{itemize}
\end{definition}

\begin{definition}[\textbf{Path}]\label{def:path}
Let $CG$ be a configurable business process graph. There is a \emph{path} $p$ between two nodes $n \in N_{CG}$ and $m \in N_{CG}$, denoted $p = n \hookrightarrow m$, if and only if (iff) there exists a sequence of nodes $n_1,\dots,n_k \in N_{CG}$ with $n=n_1$ and $m=n_k$ such that for all $i \in {1,\dots,k-1}$ holds $(n_i,n_{i+1}) \in CG$.
\end{definition}

\begin{definition}[\textbf{Mapping between the edges in $G$ and $CG$}]
Let $CG$ = $(G^\ast$, $\alpha_{G^\ast}$, $\gamma_{G^\ast}$, $\eta_{G^\ast})$ be a configurable business process graph, $N_{G^\ast}$ is the set of nodes in $CG$. $CG$ is obtained as a result of merging a set of business process variants $\mathcal{G}$. For each business process graph $G\in\mathcal{G}$, $\mathit{pid}_G\in\mathcal{I}$ is the process identifier of $G$, $N_G$ is the set of nodes in $G$, and for each node $n\in N_G$, $\lambda_G(n)$ is the label of $n$. $CG$ is obtained as a result of merging a set of business process variants, including the one represented by $G$. 
The following two functions define mappings between edges in a business process graph $G$ and a corresponding configuration business process graph $CG$. 
\begin{itemize}
	\item $\textsf{M}^\textsf{E}_\mathit{G,CG}: G \rightarrow G^\ast$ is an injective function that maps every edge in $G$ to its corresponding path in $CG$, where $\textsf{M}^\textsf{E}_\mathit{G,CG} = \{(e,p)\in G\times G^\ast \ \vert \ \forall e\in$ $G$ $\exists p\subseteq G^\ast$ such that for each $e' \in p$, ${pid}_G \in \alpha_{G^\ast}(e')\}$.
	\item $\textsf{M}^\textsf{E*}_\mathit{CG,G}: G^\ast \nrightarrow G$ is an in partial function that maps certain path in $CG$ to its corresponding edge in $G$, where $\textsf{M}^\textsf{E}_\mathit{G,CG} = \{(e,p)\in G\times G^\ast \ \vert \ \forall e\in$ $G$ $\exists p\in G^\ast$ such that for each $e' \in p$, ${pid}_G \in \alpha_{G^\ast}(e')\}$.
\end{itemize}
\end{definition}

As the foundation of our proposed co-evolution method in the article, the consolidation of the process graph into configurable process graph has actually identified all the mappings of nodes and edges between the initial process variants and configurable process models.

Given two business process graphs, $G_1$ and $G_2$ and their mapping $M_{G,CG}^N$ and $M_{G,CG}^E$ with their initial merged model $CG$, and $G_1'$ which is the new version of $G_1$, the overall co-evolution algorithm (Algorithm \ref{overall}) starts by comparing the new version of process variant $G_1'$ and the old version of process variant $G_1$ to obtain the change operation set by using the function \emph{CompareTheGraph} (c.f. Section \ref{subsubsec2.4}). Then, the function \emph{OperationJudgment} is employed to obtain the corresponding change propagation operations (Line 5). For example, the change operation "Insert Node" on $G_1$ causes the corresponding change propagation operation "Insert Node" from $G_1$ to $G_3$. Each element in $CO_1$ represents a change propagation operation to be executed from $G_1$ to $G_3$. Here, $CG$ is defined to denote the initial version of the configurable process graph (Line 6). 

\begin{algorithm}
\begin{footnotesize}
\caption{Overall Algorithm}\label{overall}
\begin{algorithmic}[1]
\State \textbf{function} Propagation(Graph $G_1$, Graph $G_1'$, Graph $G_2$, Conf.Graph $G_3$, $M_{G,CG}^N$, $M_{G,CG}^E$)

\State \textbf{init}

\State Graph $CG$, \{Change Propagation Operation\} $CO_1 \Leftarrow \phi$, \{Change Propagation Operation\} $CO_2 \Leftarrow \phi$

\State \textbf{begin}

\State $CO_1 \Leftarrow OperationJudgment( CompareTheGraph(G_1,G_2) )$
\State $CG \Leftarrow G_3$
\State $ChangePropagationG2CG(G_1,G_3, M_{G,CG}^N, M_{G,CG}^E, CO_1)$
\State $CleanGraph(G_3)$

\State $CO_2 \Leftarrow OperationJudgment( CompareTheGraph (CG,G_3) )$
\State $ChangePropagationCG2G(G_2, CG, M_{G,CG}^N, M_{G,CG}^E, CO_2)$
\State $CleanGraph(G_2)$

\State \textbf{end}
\end{algorithmic}
\end{footnotesize}
\end{algorithm}

Next, the change propagation is conducted from process variant to configurable process graph by leveraging function $ChangePropagationG2CG$ (c.f. Algorithm \ref{propagationG2CG}) (Line 7). After change propagation, the cleaning operations on $G_3$ are performed to delete the illegal parts using the function $CleanGraph$ (c.f. Algorithm \ref{cleaning_algo}) (Line 8). So far, $G_3$ is the new version of the configurable business process graph. Using the same method, we detect the changes between $G_3$ and $CG$ (Line 9) and propagate these changes to another process graph (variant) $G_2$ using function $ChangePropagationCG2G$ (c.f. Algorithm \ref{propagationCG2G}) (Line 10). Finally, $G_2$ is cleaned to remove the illegal parts on the model (Line 11).

\begin{figure*}
    \centering
    \includegraphics[width=150mm]{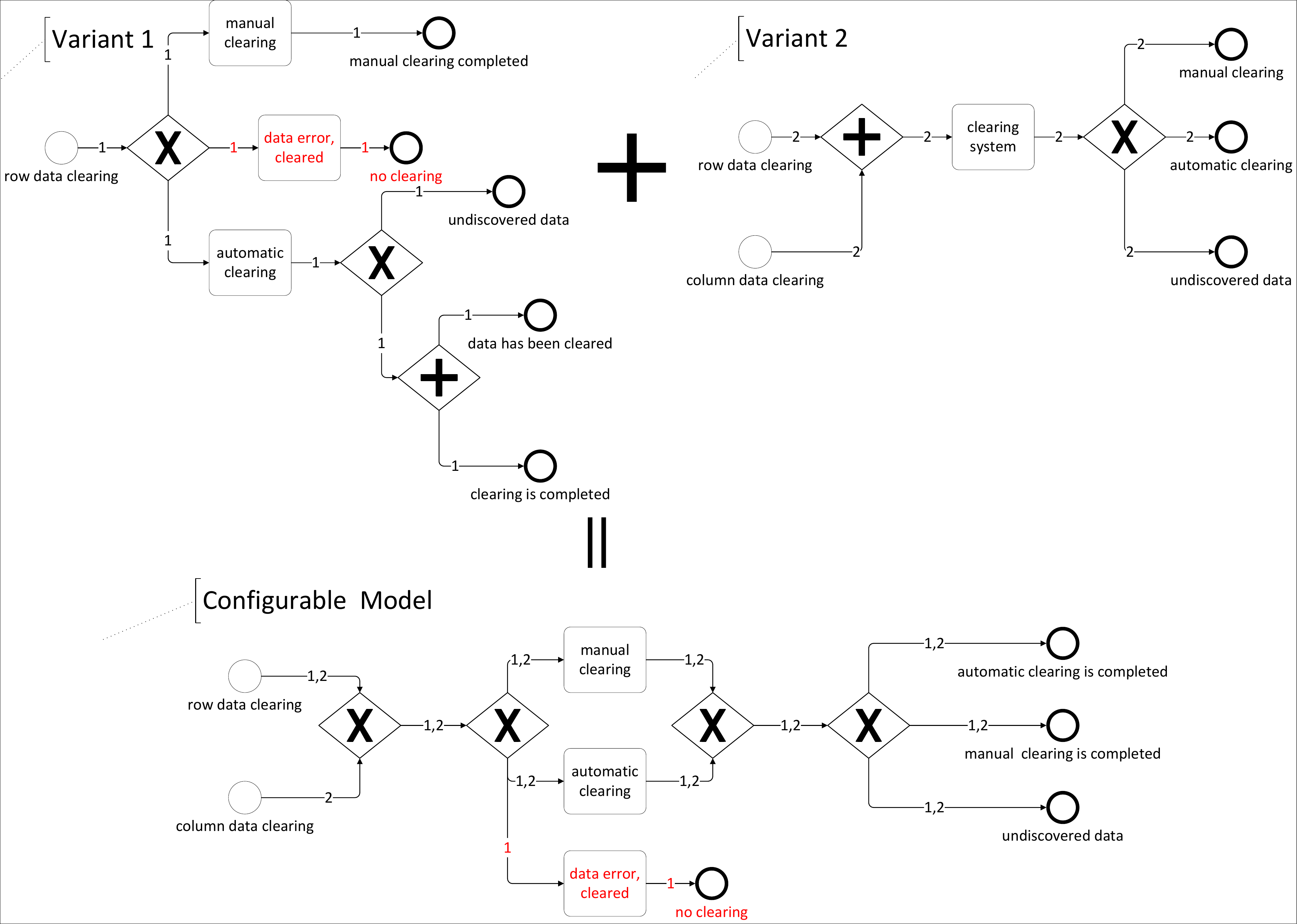}
    \caption{ Change propagation example}
    \label{fig:p2}
\end{figure*}

Considering that there may be errors in the row data that need to be deleted in process variant 1. We added a new activity, "Data error, cleared", behind the connector of process variant 1 (Fig. \ref{fig:p2}).

We can propagate the changes of Variant 1 to the configurable process model by re-merging the new version of Variant 1 and Variant 2. However, re-merging means that a merging process with high computational complexity needs to be performed again. If the changes are propagated directly to the configurable business process model, the large amount of calculation caused by re-merging will be avoided. As shown in Fig. \ref{fig:p2}, the "Data Error, cleared" activity (marked in red) has been appended behind the connector, and the merged model shows the result of the propagation (marked in red).

\begin{figure*}
    \centering
    \includegraphics[width=150mm]{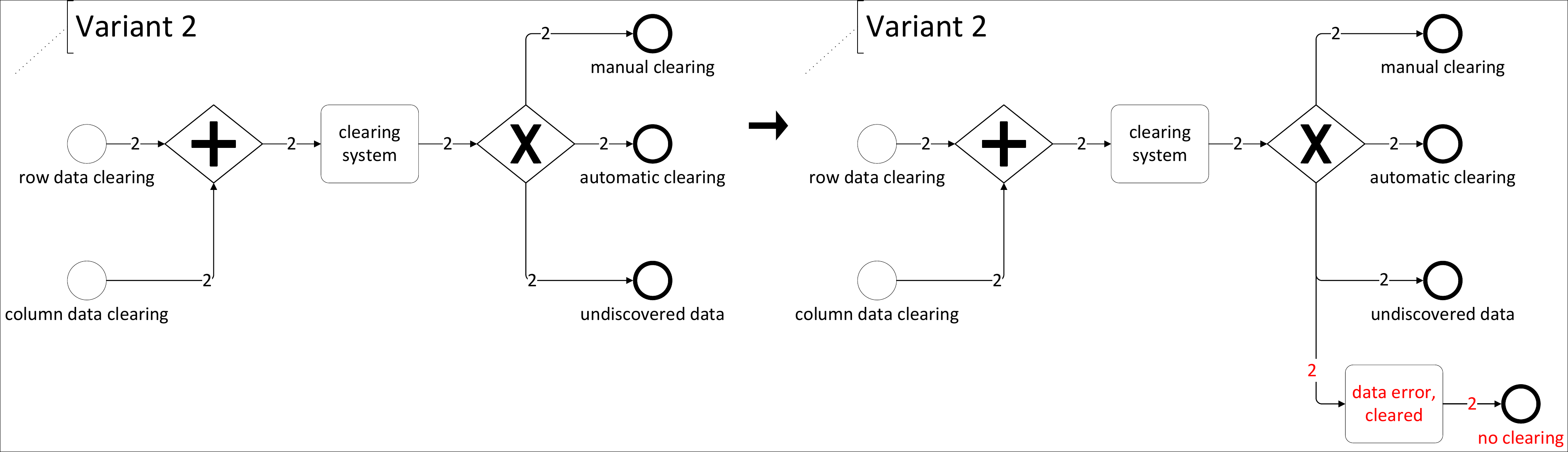}
    \caption{ Change propagation example}
    \label{fig:p2.1}
\end{figure*}

As shown in Fig. \ref{fig:p2.1}, the changes occurring in the configurable process models are also propagated to the other variants. The "Data Error, cleared" activity (marked in red) has been appended behind the OR connector of Variant 2. Hereto, the co-evolution of the process family with Variant 1 and Variant 2 is achieved.

\begin{figure*}
    \centering
    \includegraphics[width=150mm]{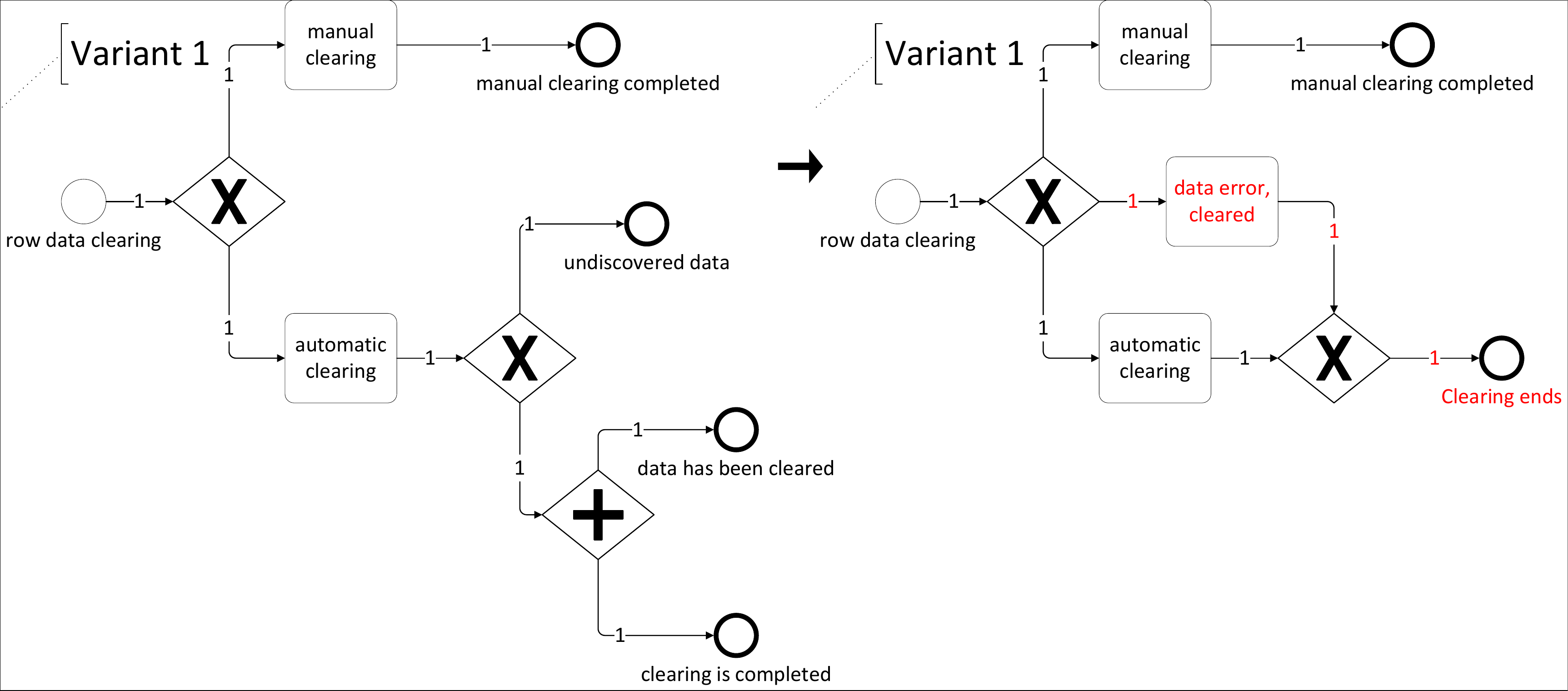}
    \caption{ Process model propagation example (Variant 1)}
    \label{fig:p4}
\end{figure*}

The overall algorithm is shown with an example. As shown in Fig.\ref{fig:p4}, the user conducts a series of modifications behind the connector of variant 1. After saving, the algorithm defines the old and new versions of variant 1, respectively as $G1$ and $G2$, and compares them to obtain two nodes to be added ("data error, cleared" and "Clearing ends"), four nodes to be deleted ("undiscovered data", "data has been cleared", "clearing is completed", and parallel connectors), three edges to be added, and multiple edges to be removed.

The above user change operation set is converted into a propagation operation set in \emph{OperationJudgment} by the following steps. First, there are four edges to be deleted, none of which belongs to "Insert Node" operation, and thus these edges to be deleted are converted into four "Delete Edge" operations. Next, the edges to be added and the nodes to be added are processed together. "Clearing ends" has one incoming edge of being added without an outgoing edge to be added. The system will use "Clearing ends" node and its edges as parameters for the "Append Node" operation. "data error, cleared" has one outgoing edge and one incoming edge, and these two edges belong to the edge to be added, but this "data error, cleared" has no edge of being deleted between the parent and child nodes, and thus the above conditions satisfy the "Add Node" operation. Finally, the nodes to be deleted become orphaned nodes, and they are removed in the cleaning operations, thus, they do not need to be propagated. In summary, user changes are converted into one "Append Node" operation, one "Add Node" operation, and four "Delete Edge" operations.

\begin{figure*}
    \centering
    \includegraphics[width=150mm]{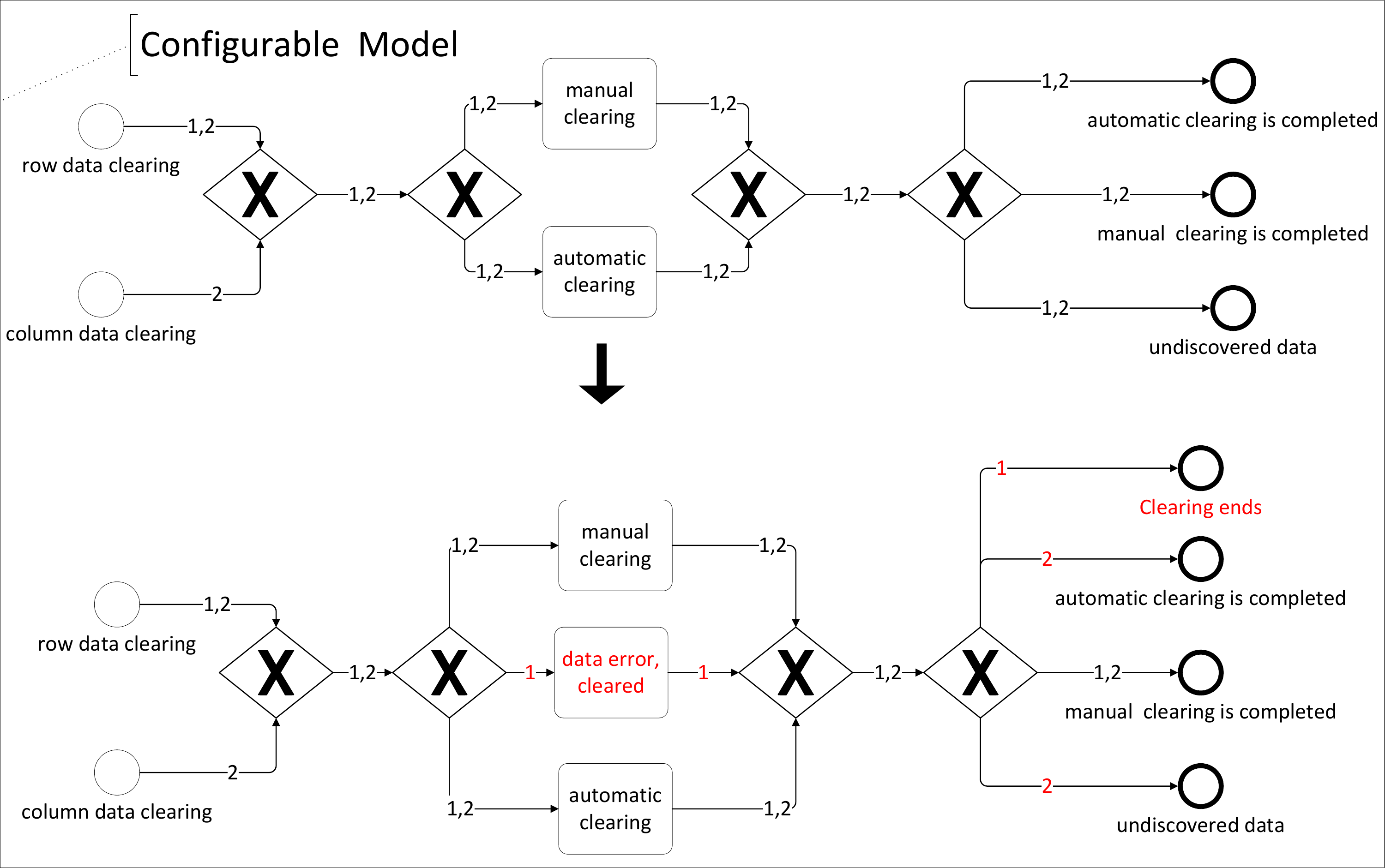}
    \caption{ Process model propagation example (Configurable model)}
    \label{fig:p5}
\end{figure*}

Further, the above-mentioned changes are propagated to $G3$ by the following steps. First, "Delete Edge" propagation operation is performed to delete the annotations of the edges corresponding to variant 1 in $G3$. Only two edges with corresponding edges in $G3$ are deleted. Next, we perform "Append Node" propagation operation on the tail connector and then conduct the "Add Node" propagation operation using "data error, cleared" and its outgoing and incoming edges as parameters. Finally, cleaning operations are performed on the propagated configurable model, and no illegal part is found in this newly generated configurable model (Fig. \ref{fig:p5}).

\begin{figure*}
    \centering
    \includegraphics[width=150mm]{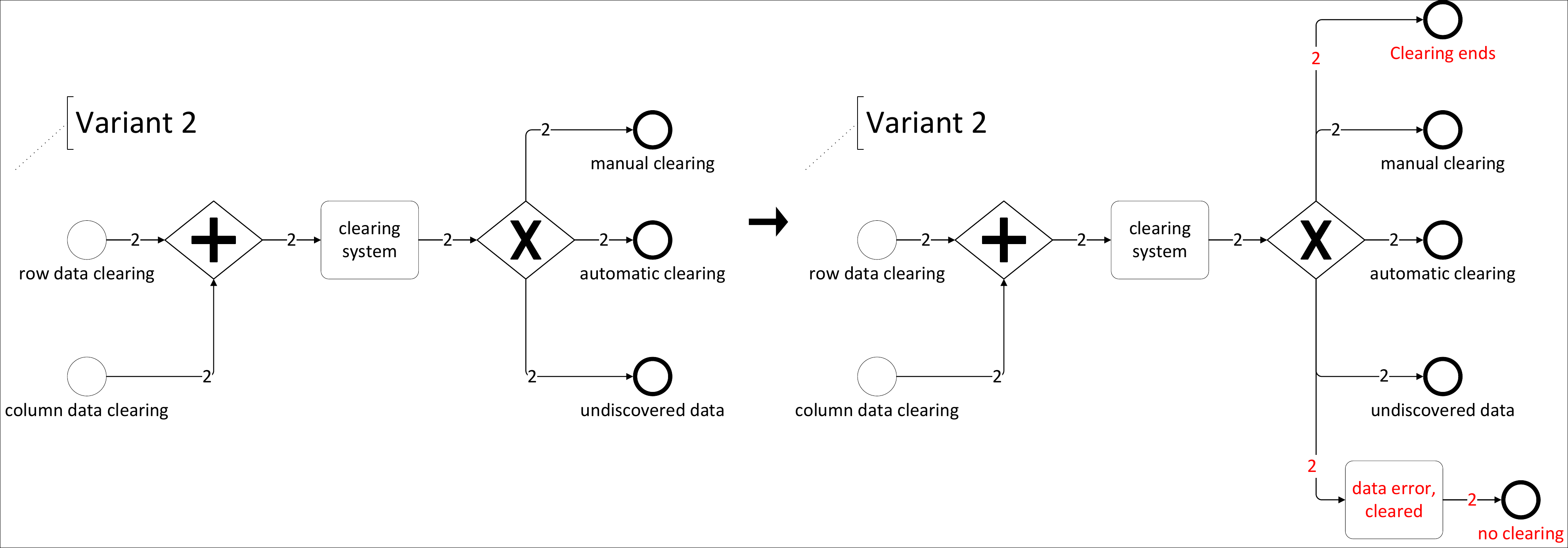}
    \caption{ Process model propagation example (Variant 2)}
    \label{fig:p5.1}
\end{figure*}

Afterward, we obtain the change operations in configurable model by the same method, and propagate them to variant 2, then perform cleaning operations on variant 2. Up to now, the entire process from modification to propagation has been completely presented, and the newly generated variant 2 is shown in Fig. \ref{fig:p5.1}.

\subsection{Propagation operations}\label{subsec3.2}

After defining the change operations of the process model, we now describe change propagating operations from process variants to configurable models and from configurable models to process variants, respectively.

\subsubsection{Propagation from variants to configurable process models}\label{subsubsec3.2.1}

\textbf{Algorithm 1} describes the change propagation from a process variant $G$ to the configurable model $CG$ when the change operation $CO$ occurs on $G$. The algorithm consists of five inputs: process variant $G$, change operation $CO$ occurring on $G$, configurable process model $CG$, node mapping \textbf{M$_{G,CG}^N$} between $G$ and $CG$, and edge mapping \textbf{M$_{G,CG}^E$}.

The mapping M$_{G,CG}^N$ refers to the mapping relationship established between the nodes of variant $G$ and the nodes of the configurable model $CG$, and the mapping M$_{G,CG}^E$ refers to that between the edges of variant $G$ and those of configurable model $CG$. The Mapping M$_{G,CG}^N$ and M$_{G,CG}^E$ save the unique ID numbers of $G$ and $CG$ so that the variants and configurable process models that are involved in change propagation can be accurately located. Node mapping M$_{G,CG}^N$ is one node-to-one node, but an edge of variant in edge mapping M$_{G,CG}^E$ may be correspond to a path in $CG$.

\begin{algorithm}
\begin{footnotesize}
\caption{Change Propagation from G2CG}\label{propagationG2CG}
\begin{algorithmic}[1]
\State \textbf{function} ChangePropagationG2CG(Graph $G$ , Conf.Graph $CG$, M$_{G,CG}^N$, M$_{G,CG}^E$, Change Operation $CO$)
\State \textbf{begin}

\If{$CO$="Insert Edge"}
\State $InsertEdge(M_{G,CG}^N(v_p), M_{G,CG}^N(v_s), CG)$
\State $\alpha_{CG}(M_{G,CG}^N(v_p), M_{G,CG}^N(v_s))$ =  $\alpha_{CG}(M_{G,CG}^N(v_p), M_{G,CG}^N(v_s))\cup \alpha_G $ 
\EndIf

\If{$CO$="Delete Edge"}
\State $\alpha_{CG}(M_{G,CG}^N(v_p), M_{G,CG}^N(v_s))$ =  $\alpha_{CG}(M_{G,CG}^N(v_p), M_{G,CG}^N(v_s))\backslash \alpha_G$ 
\EndIf

\If{$CO$="Insert Node"}
\State $v_1= M_{G,CG}^N(p(G.v)) ; v_2= M_{G,CG}^N(s(G.v))$
\If{$\vert \alpha_{(v_1,v_2)} \vert >1$}
\State $m=(c, XOR) ; n=(c, XOR)$
\State $InsertNode(v_1,v_2,m,CG)$ ; $InsertNode(m,v_2,n,CG)$
\State $v_1=m ; v_2=n$
\EndIf
\State $InsertNode(v_1,v_2,v,CG)$
\EndIf

\If{$CO$="Add Node"}
\State $v_1= M_{G,CG}^N(p(G.v))$ ; $v_2= M_{G,CG}^N(s(G.v)) $
\State $AddNode(v_1,v_2,v,CG)$
\State $\alpha_ {CG}(v_1,v) =  \alpha_ {CG}(v_1,v)\cup \alpha_ G $ ; $\alpha_ {CG}(v,v_2) =  \alpha_ {CG}(v,v_2)\cup \alpha_ G$
\EndIf

\If{$CO$="Append Node"}
\State $v_1= M_{G,CG}^N(v_p) ; v_2=v_1$
\If{$\tau (v_p)\in{e,f}\quad and \quad \vert \alpha_{(p(v_1),v_1)} \vert >1$}
\State $m=(c, XOR)$
\State $InsertNode(p(v_1),v_1,m,CG)$ ; $AppendNode(m,v_2,CG)$
\State $\alpha_{CG}(m, v_2)=\alpha_{CG}(m, v_2)\backslash\alpha_G$
\EndIf
\State $AppendNode(v_1,v_s,CG)$
\EndIf

\If{$CO$="Change Node Annotation"}
\State $v_1= M_{G,CG}^N(CO.v)$
\If{$ \vert \alpha_{CG}(p(v_1),v_1) \vert =1 \quad or\quad \alpha_{CG}(v_1,s(v_1)) \vert =1 $}
\State $\alpha_{CG}(v_1)=\alpha_{G}(v)$
\EndIf

\If{$ \vert \alpha_{CG}(p(v_1),v_1) \vert >1\quad and \quad\ s(v_1)=\emptyset $}
\State $m=(c, XOR)$
\State $InsertNode(p(v_1), v_1, m, CG)$ ; $AppendNode(m, G.v, CG)$
\State $\alpha_{CG}(m, v_1)=\alpha_{CG}(m, v_1)\backslash\alpha_G$
\EndIf

\If{$ \vert \alpha_{(p(v_1),v_1)} \vert >1\quad and \quad\vert\alpha_{(v_1,s(v_1))} \vert >1 $}
\State $m=(c, XOR) ; n=(c, XOR)$
\State $InsertNode(p(v_1), v_1, m, CG)$ ; $InsertNode(v_1,s(v_1),n,CG)$
\State $AddNode(m,n,G.v,CG)$
\State $\alpha_{CG}(m, v_1)=\alpha_{CG}(m, v_1)\backslash\alpha_G$ ; $\alpha_{CG}(v_1, n)=\alpha_{CG}(v_1, n)\backslash\alpha_G$
\EndIf
\EndIf

\State $M_{G,CG}^N = M_{G,CG}^N \cup N_{new}$ ; $M_{G,CG}^N = M_{G,CG}^N \backslash N_{del}$
\State $M_{G,CG}^E = M_{G,CG}^E \cup E_{new}$ ; $M_{G,CG}^E = M_{G,CG}^E \backslash E_{del}$

\State \textbf{end}
\end{algorithmic}
\end{footnotesize}
\end{algorithm} 

The propagation operations to be performed is determined by the change operation $CO$. Algorithm 1 defines 6 change propagation operations:

\begin{enumerate}

\item \textbf{Insert Edge} (Line 3-6): The "Insert Edge" change operation is performed on $G$ in this case. If an edge is inserted between $v_p$ and $v_s$ on $G$ and there is an existing edge that corresponds to the inserted edge on $G$ before propagation, the annotation $\alpha_G$ will be inserted onto the edge, which means performing the "Insert Edge Annotation" change operation in $CG$. Otherwise, a new edge is inserted between the corresponding nodes of $v_p$ and $v_s$ the $CG$, which means performing an "Insert Edge" operation.
\item \textbf{Delete Edge} (Line 7-9): The "Delete Edge" change operation is performed on $G$ in this case. The annotation $\alpha_G$ on the corresponding edge of $CG$ is cleared through the operation "Delete Edge Annotation". If the annotation is empty after deletion, the operation "Delete Edge" will be performed. 
\item \textbf{Insert Node} (Line 10-18): The "Insert Node" operation to insert node $v$ is performed on $G$ in this case. Before the change is propagated from $G$ to $CG$, we identify $v_1$ and $v_2$ in $CG$ corresponding to the parent node and child node of $v$ in $G$, respectively. If the annotation of the edge between $v_1$ and $v_2$ is more than 1, we will insert two $XOR$ connectors, $m$ and $n$, between $v_1$ and $v_2$ and insert a new node $v$ between the connectors. If this annotation is $\leq$ 1, we insert $v$ directly on the path. In each case, the edge annotation needs to be modified.
\item \textbf{Add Node} (Line 19-23): The "Add Node" operation to add node $v$ is performed on $G$ in this case. To this end, we first identify the nodes $v_1$ and $v_2$ in $CG$ corresponding to the parent node and child node of $v$ in $G$, respectively, and then add a path with only node $v$ between $v_1$ and $v_2$. Finally, we add an edge annotation $\alpha_G$ on the path on $CG$.
\item \textbf{Append Node} (Line 24-32): The "Append Node" operation is performed on $G$ in this case. During propagation, we append a node behind $v_1$ that is the corresponding node $v_p$ of the $G$. If the type of $v_p$ is "event" or "function", and the number of annotation of its incoming edge $(p(v_1),v_1)$ is more than 1, and we will add a connector $m$ in front of the corresponding node $v_1$ and insert a copy $v_2$ of the corresponding node $v_1$ behind the connector $m$, and then append a node $v_s$ behind the copy node. Since the "Prepend Node" operation is similar to "Append Node", and thus it is omitted from the algorithm.
\item \textbf{Change Node Annotation} (Line 33-49): The "Change Node Annotation" operation is performed on $G$ in this case. During propagation, this operation in $CG$ needs to be performed in multiple cases: If the number of annotations on the incoming edge $(p(v_1),v_1)$ or outgoing edge $(v_1,s(v_1)$ of the node $v_1$ is equal to 1, we will simply perform the operation "Change Node Annotation" on $CG$ (Line 35-37); If the number of annotation of the incoming edge $(p(v_1),v_1)$ is \textgreater 1 and there is no outgoing edge, we will insert an $XOR$ connector $m$ before the node $v_1$ and delete the annotation $\alpha_G$ about $G$ on edge, then append the node $G.v$ whose annotation has been updated in $G$ behind the connector $m$ (Line 38-42). If the number of annotations of both incoming edge $(p(v_1),v_1)$ and outgoing edge $(v_1,s(v_1))$ are more than 1, we will add $XOR$ connectors $m$ and $n$ before and after the node, and then insert a new path containing only the node whose annotation has been updated in $G$ between the two connectors $m$ and $n$, and finally, we delete the related edge annotation $\alpha_G$ from the edge $(m, v_1)$ and $(v_1, n)$ (Line 43-48). Note that if the annotation of the outgoing edge is more than 1 and there is no incoming edge, the opposite operation will be performed (namely, we insert a connector behind the node and prepend the new node in front of the connector), which is omitted in \textbf{Algorithm 1}. 

\end{enumerate}

After the propagation is complete, we further update the node and edge mapping between the variant and the configurable model to ensure their correctness(Line 50-51): Add the new node mapping to M$_{G,CG}^N$, delete the node mapping which should be deleted from M$_{G,CG}^N$. Add the new edge mapping to M$_{G,CG}^E$, delete the edge mapping, which should be deleted from M$_{G,CG}^E$.

\subsubsection{Propagation from configurable process models to variants}\label{subsubsec3.2.2}

\textbf{Algorithm 2} describes the propagation of change operation $CO$ from configurable process model $CG$ to the relevant variant $G$. The algorithm consists of five inputs: Mapping \textbf{var} saving two variants and their IDs, configurable model \textbf{CG}, change operation \textbf{CO} occurring in $CG$, node mapping \textbf{M$_{G,CG}^N$}, and edge mapping \textbf{M$_{G,CG}^E$}. 

\begin{algorithm}
\begin{footnotesize}
\caption{Change Propagation from CG2G}\label{propagationCG2G}
\begin{algorithmic}[1]
\State \textbf{function} ChangePropagationCG2G(Map$<ID,G> var$, Conf.Graph $CG$, M$_{G,CG}^N$, M$_{G,CG}^E$, Change Operation $CO$)
\State \textbf{begin}

\If{$CO$="Insert Edge"}
\While{$\alpha_G \in \alpha_{CG}(v_p,v_s)$}
\State $InsertEdge(M_{G,CG}^N(v_p),  M_{G,CG}^N(v_s),G)$
\EndWhile
\EndIf

\If{$CO$="Delete Edge"}
\While{$\alpha_G \in \alpha_{CG}(v_p,v_s)$}
\State $DeleteEdge(M_{G,CG}^N(v_p),  M_{G,CG}^N(v_s),G)$
\EndWhile
\EndIf

\If{$CO$="Modify Node Annotation"}
\While{$\alpha_ G \in \alpha_ {CG}(p(CO.v),CO.v)$}
\State $\alpha_ G(M_{G,CG}^N(CO.v))=\alpha_ {CG}(CO.v)$
\EndWhile
\EndIf

\If{$CO$="Append Node"}
\While{$\alpha_ G \in \alpha_ {CG}(p(CO.v),CO.v)$}
\State $AppendNode(M_{G,CG}^N(p(CO.v)), CO.v ,G)$
\EndWhile
\EndIf

\If{$CO$="Add Node"}
\While{$\alpha_ G \in \alpha_ {CG}(p(CO.v),CO.v)$}
\State $v_1= M_{G,CG}^N(p(CO.v))$ ; $v_2= M_{G,CG}^N(s(CO.v))$
\If{$v_1=\emptyset \quad or \quad v_2=\emptyset$}
\State $v_1=CreatAuxNode(p(CO.v), G, CG)$ 
\State $v_2=CreatAuxNode(s(CO.v), G, CG)$
\EndIf
\State $AddNode(v_1,v_2,CO.v,G)$
\EndWhile
\EndIf

\If{$CO$="Insert Node"}
\While{$\alpha_ G \in \alpha_ {CG}(p(CO.v),CO.v)$}
\State $v_1= M_{G,CG}^N(p(CO.v))$ ; $v_2= M_{G,CG}^N(s(CO.v))$
\If{$v_1=\emptyset \quad or \quad v_2=\emptyset$}
\State $v_1=CreatAuxNode(p(CO.v), G, CG)$ 
\State $v_2=CreatAuxNode(s(CO.v), G, CG)$
\EndIf
\State $InsertNode(v_1,v_2,CO.v,G)$
\EndWhile
\EndIf

\If{$CO$="Insert Edge Annotation"}
\State $\alpha_ {Label} = New.\alpha_ {CG}(CO.e) \backslash \alpha_ {CG}(CO.e)$ 
\While{$\alpha_ G \in \alpha_ {Label}$}
\State $v_1= M_{G,CG}^N(v_p)$ ; $v_2= M_{G,CG}^N(v_s)$ 
\If{$v_1=\emptyset \quad or \quad v_2=\emptyset$}
\State $v_1=CreatAuxNode(p(CO.v), G, CG)$ 
\State $v_2=CreatAuxNode(s(CO.v), G, CG)$
\EndIf
\State $InsertEdge(v_1, v_2,G)$ 
\EndWhile
\EndIf

\If{$CO$="Delete Edge Annotation"}
\State $\alpha_ {Label} = \alpha_ {CG}(CO.e) \backslash New.\alpha_ {CG}(CO.e)$ 
\While{$\alpha_ G \in \alpha_ {Label}$}
\State $v_1= M_{G,CG}^N(v_p)$ ; $v_2= M_{G,CG}^N(v_s)$ 
\If{$v_1=\emptyset \quad or \quad v_2=\emptyset$}
\State $v_1=CreatAuxNode(p(CO.v), G, CG)$ 
\State $v_2=CreatAuxNode(s(CO.v), G, CG)$
\EndIf
\State $DeleteEdge(v_1, v_2,G)$ 
\EndWhile
\EndIf

\State $M_{G,CG}^N = M_{G,CG}^N \cup N_{new}$ ; $M_{G,CG}^N = M_{G,CG}^N \backslash N_{del}$
\State $M_{G,CG}^E = M_{G,CG}^E \cup E_{new}$ ; $M_{G,CG}^E = M_{G,CG}^E \backslash E_{del}$

\State \textbf{end}
\end{algorithmic}
\end{footnotesize}
\end{algorithm} 

Totally 8 propagation operations are involved in Algorithm 2. 

\begin{enumerate}

\item \textbf{Insert Edge} (Line 3-7): The "Insert Edge" operation is performed on $CG$ in this case. If an edge is inserted from $v_p$ to $v_s$ with the annotation $\alpha_{CG}$ on $CG$, We perform the "Insert Edge" operation from the corresponding starting node of $v_p$ and target node $v_s$ on the corresponding process variant $G$ whose process identifier is in the set of $\alpha_{CG}$.
\item \textbf{Delete Edge} (Line 8-12): The "Delete Edge" operation is performed on $CG$ in this case. If an edge from $v_p$ to $v_s$ is removed from $CG$ with the edge annotation $\alpha_{CG}$. We conduct the "Delete Edge" operation on the corresponding variant $G$. To this end, we identify the edge in the variant $G$ corresponding to the deleted edge in the $CG$, and delete it.
\item \textbf{Modify Node Annotation} (Line 13-17): The "Modify Node Annotation" operation is performed on $CG$ in this case. During propagation, we detect the node whose annotation has been updated in $CG$, and then further identify the corresponding node of the above-detected node in each variant according to the mapping $M_{G,CG}^N$. Afterward, we perform the "Modify Node Annotation" operation on each corresponding node.
\item \textbf{Append Node} (Line 18-22): The "Append Node" operation is performed on $CG$ in this case. During propagation, we detect the variant modified according to the annotation on the incoming edge of the appended node in $CG$, and then perform the "Append Node" operation at the corresponding position of the variant. Since the "Prepend Node" operation is similar to "Append Node", "Prepend Node" has been omitted from the algorithm.
\item \textbf{Add Node} (Line 23-32): The "Add Node" operation is performed on $CG$ in this case. During propagation, we identify the corresponding variant according to the annotations on the incoming and outgoing edges of $v$, and then perform the "Add Node" operation on the identified variant. Further, we identify nodes $v_1$ and $v_2$ in $G$ respectively corresponding to the parent node and child node of $v$ in $CG$, insert a path containing only $v$ between the two nodes, and add the edge annotation. If $v_1$ or $v_2$ cannot be found in the variant, a connector will be inserted at the corresponding position of the variant to serve as $v_1$ or $v_2$.
\item \textbf{Insert Node} (Line 33-42): The "Insert Node" operation is performed on $CG$ to insert node $v$ in this case. During propagation, we identify the corresponding variant according to the annotations on the incoming and outgoing edges of $v$, and then perform the "Insert Node" operation on the identified variant. We further identify nodes $v_1$ and $v_2$ in $G$, respectively corresponding to the parent node and child node of $v$ in $CG$, and insert $v$ on the path. If $v_1$ or $v_2$ cannot be found in the variant, which means that the parent (child) node of $v$ is an auxiliary node, a connector will be inserted at the corresponding position in the variant to act as $v_1$ or $v_2$.
\item \textbf{Insert Edge Annotation} (Line 43-53): The "Insert Edge Annotation" operation is performed on $CG$ to add the new annotation on edge $e$ in this case. Based on the added annotation, we detect the corresponding variant during propagation and then identify the nodes in the variant corresponding to the start node and end node of edge $e$. Further, we perform the "Insert Edge" operation. If no corresponding node can be found in the variant, the corresponding node in $CG$ will be copied into the variant.
\item \textbf{Delete Edge Annotation} (Line 54-64): The "Delete Edge Annotation" operation is performed on $CG$ to delete some annotations on edge $e$ in this case. According to the deleted annotation, we detect the corresponding variant during propagation. Subsequently, we identify the nodes in the variant corresponding to the start node and end node of edge $e$ and perform the "Delete Edge" operation. If no corresponding node can be found in the variant, the corresponding node in $CG$ will be copied into the variant. If "Delete Edge" operation removes all the annotations of edge $e$, edge $e$ will be removed through the cleaning operations.
\end{enumerate}

After the propagation is complete, we further update the mapping(Line 65-66): Add the new node mapping to M$_{G,CG}^N$, delete the node mapping, which should be deleted from M$_{G,CG}^N$. Add the new edge mapping to M$_{G,CG}^E$, delete the edge mapping, which should be deleted from M$_{G,CG}^E$.

\subsection{Cleaning operations}\label{subsec3.3}

After finishing the change propagation between variants and configurable models, the newly obtained process model may have some \emph{illegal} parts caused by the modification of the propagation operation, and thus this new model may not conform to the EPC or BPMN syntax. It is necessary to remove these illegal parts on $G$ or $CG$ to ensure the correctness of process models in a specified language. To this end, we propose \emph{cleaning} operations to clean up illegal parts of the post-propagation process model. The cleaning operations of the configurable process model $CG$ are shown in Algorithm \ref{cleaning_algo}. The cleaning operations on process variants are similar to the operations defined in this algorithm.

\begin{algorithm}
\begin{small}
\caption{Clean Operation}\label{cleaning_algo}
\begin{algorithmic}[1]
\State \textbf{function} CleanGraph(Conf.Graph $CG$)
\State \textbf{begin}

\While{$\tau (N_{CG.v})\in {c}$}

\If{$\vert p(v) \vert > 1 \quad and \quad \vert s(v) \vert > 1$}
\State Create a new node $v_1$ and make $v_1.type=v.type$
\State $InsertEdge(v_1,s(v),CG)$
\State $DeleteEdge(v, s(v),CG)$
\State $InsertEdge(v,v_1,CG)$
\EndIf

\If{$\vert p(v) \vert < 1 \quad and \quad \vert s(v) \vert < 1$}
\State $InsertEdge(p(v),s(v),CG)$
\State $\alpha_{CG}(p(v),s(v)) = \alpha_{CG}(v,s(v))$
\State $DeleteEdge(v, s(v),CG)$
\State $DeleteEdge(p(v), v, CG)$
\State $RemoveNode(v)$
\EndIf

\If{$s(v).type = v.type$}
\While{$s(s(v))$}
\State $InsertEdge(v, s(s(v)), CG)$
\State $\alpha_ {CG}(v,s(s(v))) = \alpha_ {CG}(s(v),s(s(v)))$
\State $DeleteEdge(s(v), s(s(v)), CG)$
\EndWhile
\State $DeleteEdge(v, s(v),CG)$
\State $RemoveNode(s(v))$
\EndIf

\EndWhile

\If{$\alpha_ {(E_{CG.e})} = \emptyset$}
\State $DeleteEdge(e.v_p , e.v_s , CG)$
\EndIf

\If{$p(N_{CG.v}) = \emptyset \quad and \quad s(N_{CG.v}) = \emptyset$}
\State $RemoveNode(v)$
\EndIf

\State \textbf{end}
\end{algorithmic}
\end{small}
\end{algorithm} 

As shown in Algorithm \ref{cleaning_algo}, the only input of the algorithm is a configurable process graph $CG$ to be cleaned. 5 illegal cases in $CG$ will be inspected and then corrected. 

\begin{itemize}
\item If a connector $v$ has multiple child nodes and parent nodes, we will create a connector $v_1$ with the same type as connector $v$ (Line 4-5), insert the edges between $v_1$ and all child nodes of $v$, and then delete all the edges of $v$ and its child nodes (Line 6-7). Finally, a new edge is inserted between $v$ and $v_1$ (Line 8).
\item If neither the number of child nodes nor that of parent nodes of a connector $v$ are more than 1, which means that the connector $v$ is a \emph{redundant} connector. This redundant connector and its edges are deleted, and then the parent and child nodes of the deleted connector are connected.
\item If the child nodes of connector $v$ still contain a connector $s(v)$, and $v$ and $s(v)$ are of the same type, it means that $v$ and $s(v)$ form continuous connectors. In this case, we will change all the child nodes of $s(v)$ into the child nodes of $v$, and delete $s(v)$ and its edges. 
\item If the annotation of an edge is empty, we will delete this edge. But deletion may cause more illegal situations.
\item If a node $v$ has neither an outgoing edge nor an incoming edge, this node is called an \emph{orphan} node, and this node will be deleted from $G$. Alternatively, if the annotation of a node $v$ is empty, this node will also be deleted.
\end{itemize}

Since the change operation on a connector $c$ may cause the ripple change operations of the neighboring edges of $c$, the mapping relationships $M_{G,CG}^N$ and $M_{G,CG}^E$ between configurable process model $CG$ and process variants $G$s need to be updated after the cleaning work is finished, which differs from the variant’s cleaning operations requiring no updating. We define execution \emph{flag}s for five illegal cases at runtime. When a case is executed, the flag will be set as \textbf{true}, indicating that the graph has been partially modified in the loop, which may lead to new illegal parts. The loop will be stopped only if all five flags are \textbf{false}.

\begin{figure*}
    \centering
    \includegraphics[width=170mm]{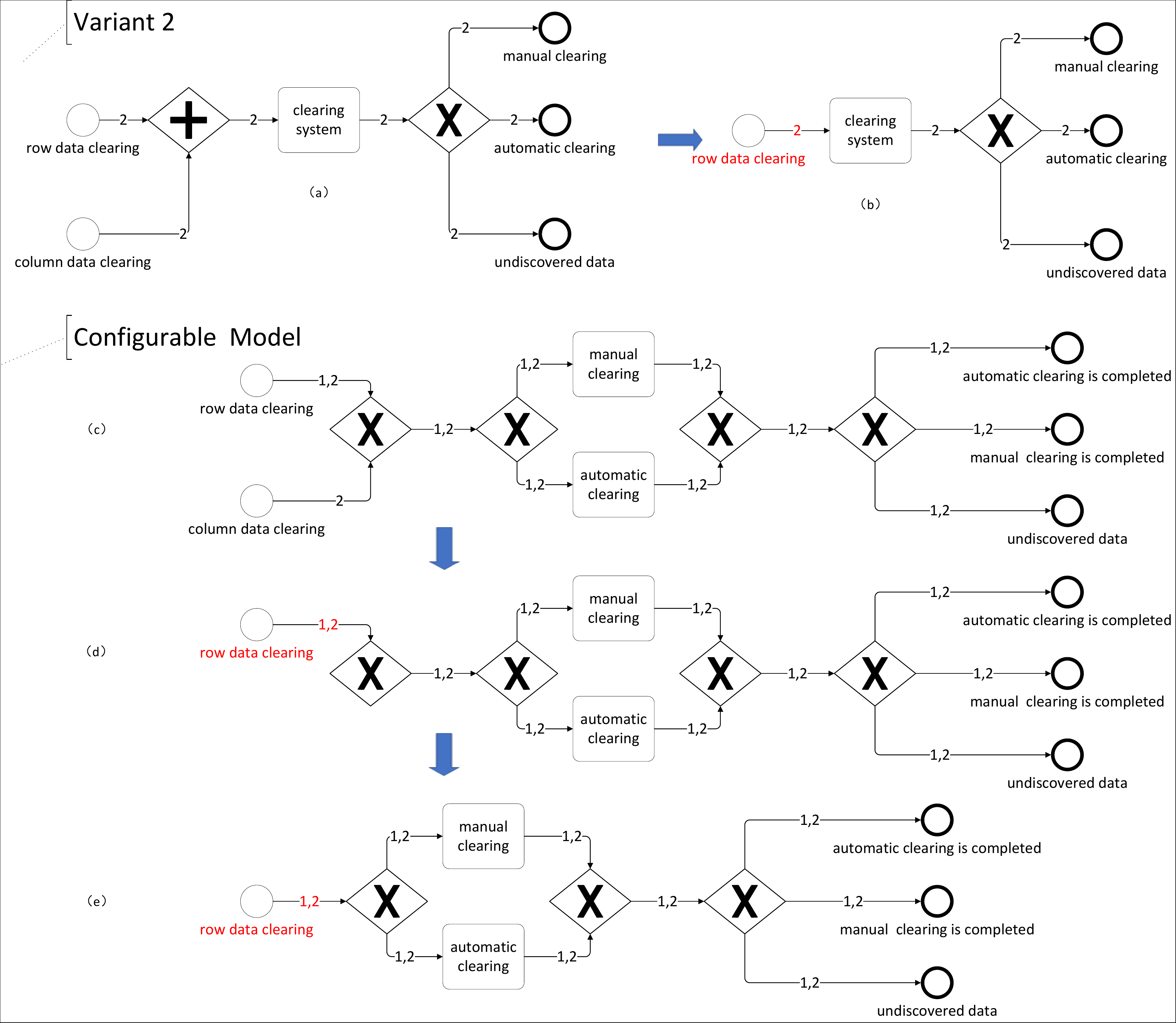}
    \caption{ Cleaning example}
    \label{fig:p3}
\end{figure*}

Figure \ref{fig:p3} demonstrates an example of applied cleaning operations after change propagates from process variant graph $G$ to $CG$. As shown in Fig.\ref{fig:p3}  (a) and (b), supposing that the events "column data clearing" and its subsequent \emph{parallel} connector are removed from Variant 2. In this case, the event of "column data clearing" and its related edges will be deleted from the corresponding configurable model. The initial propagation results are shown in Fig.\ref{fig:p3} (c). After the propagation is completed, the subsequent connector of the activity "row data clearing" has only one incoming edge and one outgoing edge, which is shown in Fig.\ref{fig:p3} (d), indicating this connector is redundant. Thus, as shown in Fig.\ref{fig:p3} (e), the cleaning operation will remove this redundant connector and directly connect the event "row data clearing" to the next connector of this redundant connector.

\section{Prototype}\label{sec4}

The co-evolution algorithm has been implemented as a tool, namely \emph{BPCE}, that is freely available on GitHub \footnote{https://github.com/Zaiwen/bp\_change\_propagation}. The tool accepts two (or more) EPCs/BPMNs represented and their configurable process model in the EMPL/BPMN format. Users can select a process variant from a list of process families and make the change(s) on the variant. The change(s) will be propagated to the configurable process model, which can be further simplified by applying the cleaning rules. After the user has reviewed and validated the resulting evolved model, the changes on the configurable process model are reversely propagated into another process variant(s). The implementation of the algorithm has also been integrated into the AProMoRe platform - a process model repository toolset (see: http://www.apromore.org).

\begin{figure*}
    \centering
    \includegraphics[width=150mm]{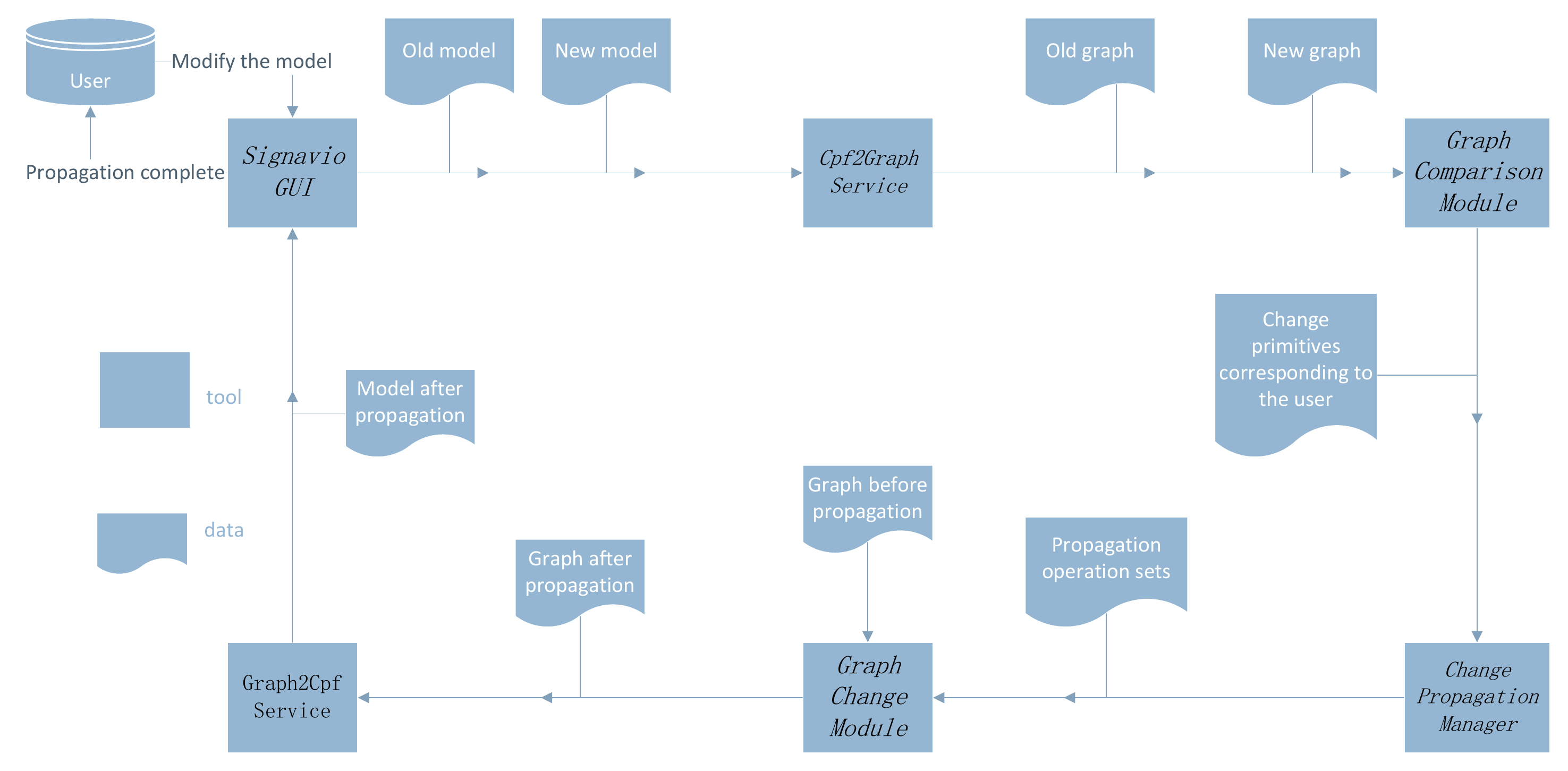}
    \caption{ System architecture of change propagation}
    \label{fig:p6}
\end{figure*}

The framework of the tool is shown in Fig. \ref{fig:p6}. This framework consists of 6 main components: \emph{Signavio GUI}, \emph{Cpf2Graph Service}, \emph{Graph Comparison Module}, \emph{Change Propagation Manager}, \emph{Graph Change Module}, and \emph{Graph2Cpf Service}. Users modify the model using \emph{Signavio GUI} and issue a change propagation request, and the system will pass the modified model and the original model together to \emph{Cpf2Graph Service} to obtain the directed graph corresponding to the model. Then, \emph{Graph Comparison Module} compares the old and new graphs to obtain the user’s modifications and their corresponding change primitives. Through \emph{Change Propagation Manager}, we determine the propagation operations to be performed, based on which \emph{Graph Change Module} modifies the correlation graph. Finally, using \emph{Graph2Cpf Service}, the directed graph is converted into a configurable business process model which is presented to the user through \emph{Signavio GUI}.

\section{Evaluation}\label{sec5}

To demonstrate and evaluate the feasibility and efficiency of our method, we conduct experiments with the SAP reference model and data from 155 commercial BPEL instance models using the implementation of the algorithm. 

For these tests, we took the SAP reference model, consisting of 604 EPCs, and constructed EPCs business process families from among them. These original SAP EPC reference models are not configurable, and they need to be classified according to \emph{matching score} of similarity. Therefore, we used a graph similarity calculation method \cite{r1} to compute the matching score \cite{dijkman2011}, obtained the similarity between graphs, and then employed the agglomerative hierarchical clustering (AHC) \cite{bouguettaya2015} algorithm to cluster these graphs. After clustering, a total of 40 process clusters are obtained with 15 process models on average in each cluster. We discarded 30 unclustered EPC models, selected one pair of models from some remaining clusters, and obtain a total of 20 pairs of models as variants. Next, we use the model merging method proposed in the literature  \cite{r1} to merge model pairs and obtain 20 business process families, with each family composed of a configurable business process model and their corresponding process variants.

Meanwhile, we took 155 BPEL models, extracted their control flows, transformed the part of control flows into BPMN graphs, and then performed hierarchical clustering operations on the BPMN graphs based on matching scores. A total of 20 process clusters were obtained from 90 BPMN graphs. Only 4 meaningful clusters were obtained from BPMN graphs due to graph over-simplicity. We selected 20 pairs of models from these 4 clusters as variants and merged them into 20 business process families consisting of configurable business process models and their corresponding variants. 

In summary, we obtained a total of 40 families of business process variants in the form of EPML or BPEL, then conducted change propagation experiments on these 40 datasets.

\subsection{Effectiveness of propagation algorithm}\label{subsec5.1}

Ensuring \emph{correctness} is a key factor affecting the effectiveness of change propagation algorithms. It is thus desirable that models are as correct as possible after changes are propagated from process variants to configurable process graphs, or from configurable process graphs to process variants.

We conducted tests with 6 change propagation operations defined in $ChangePropagationG2CG$ (c.f. Algorithm \ref{propagationG2CG}) and 8 change propagation operations described in $ChangePropagationCG2G$ (c.f. Algorithm \ref{propagationCG2G}) for each data set to verify the correctness of models after change propagation. In addition to the 6 operations listed in Algorithm \ref{propagationG2CG}, there was also a "Prepend Node" operation to be tested. We omitted the "Insert Edge" change propagation operation since it is included in other operations. Likewise, we omit "Insert Edge" from change propagation operations listed in Algorithm \ref{propagationCG2G}. Basically, these propagation operations can cover almost all the changes between variants and configurable models, and they conform to BPMN rules \cite{r14}.

\begin{table}[!htbp]
\caption{\centering The accuracy of change propagation (variants to configurable process model)}\label{tab1}
\resizebox{83mm}{15mm}{
\begin{tabular}{|c|c|c|c|}
\hline
Operation & Correct Rate $(\%)$ & Operation & Correct Rate $(\%)$ \\
\hline
Delete Edge & 100 & Delete Edge & 100 \\
Append Node & 94.7 & Append Node & 100 \\
Prepend Node & 95 & Prepend Node & 100 \\
Add Node & 94.4 & Add Node& 100 \\
Insert Node & 95 & Insert Node & 100 \\
Modify Node Annotation & 100 & Modify Node Annotation & 100 \\
\hline
\end{tabular}}
\begin{tablenotes}
\item The left part is the operation on the EPC set, and the right part is the operation on the BPEL set
\end{tablenotes}
\end{table}

\begin{table}[!htbp]
\caption{\centering The accuracy of change propagation (from configurable process model to variants)}\label{tab2}
\resizebox{83mm}{18mm}{
\begin{tabular}{|c|c|c|c|}
\hline
Operation & Correct Rate $(\%)$ & Operation & Correct Rate $(\%)$ \\
\hline
Delete Edge & 100 & Delete Edge & 100 \\
Append Node & 100 & Append Node & 100 \\
Prepend Node & 100 & Prepend Node & 100 \\
Add Node & 100 & Add Node& 100 \\
Insert Node & 100 & Insert Node & 100 \\
Modify Node Annotation & 100 & Modify Node Annotation & 100 \\
Insert Edge Annotation & 100 & Insert Edge Annotation & 100 \\
Delete Edge Annotation & 100 & Delete Edge Annotation & 100 \\
\hline
\end{tabular}}
\begin{tablenotes}
\item The left part is the operation on the EPC set, and the right part is the operation on the BPEL set
\end{tablenotes}
\end{table}

Table \ref{tab1} and Table \ref{tab2} summarize the result for the accuracy of change propagation with Algorithm \ref{propagationG2CG} and \ref{propagationCG2G}, respectively. Most propagation operations exhibit an average correctness rate of more than 90\% on 40 model sets. The BPEL model is relatively simple with a high average correctness rate, while the EPC model is more complex with a low correctness rate, which is still higher than 93\%. Therefore, it could be concluded that the propagation operations are feasible and correct, and that they are applying to the vast majority of model modifications.

\subsection{Efficiency evaluation}\label{subsec5.2}

We also conducted tests with 40 model sets to assess the efficiency of the change propagation operators.

\begin{figure}
    \centering
    \includegraphics[width=80mm]{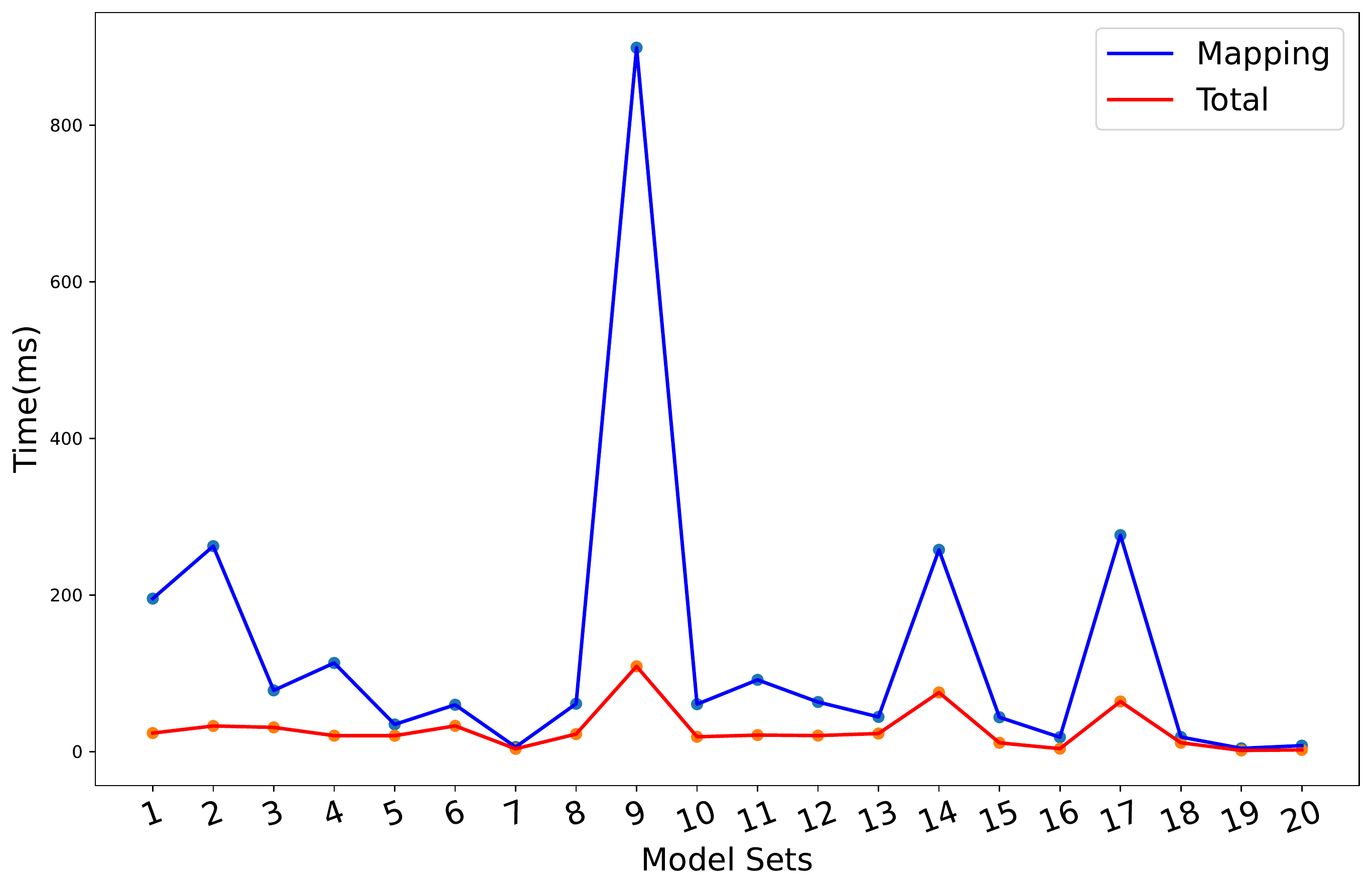}
    \caption{\centering Time-consuming comparison (EPC)}
    \label{time_comparision_EPC}
\end{figure}

\begin{figure}
    \centering
    \includegraphics[width=80mm]{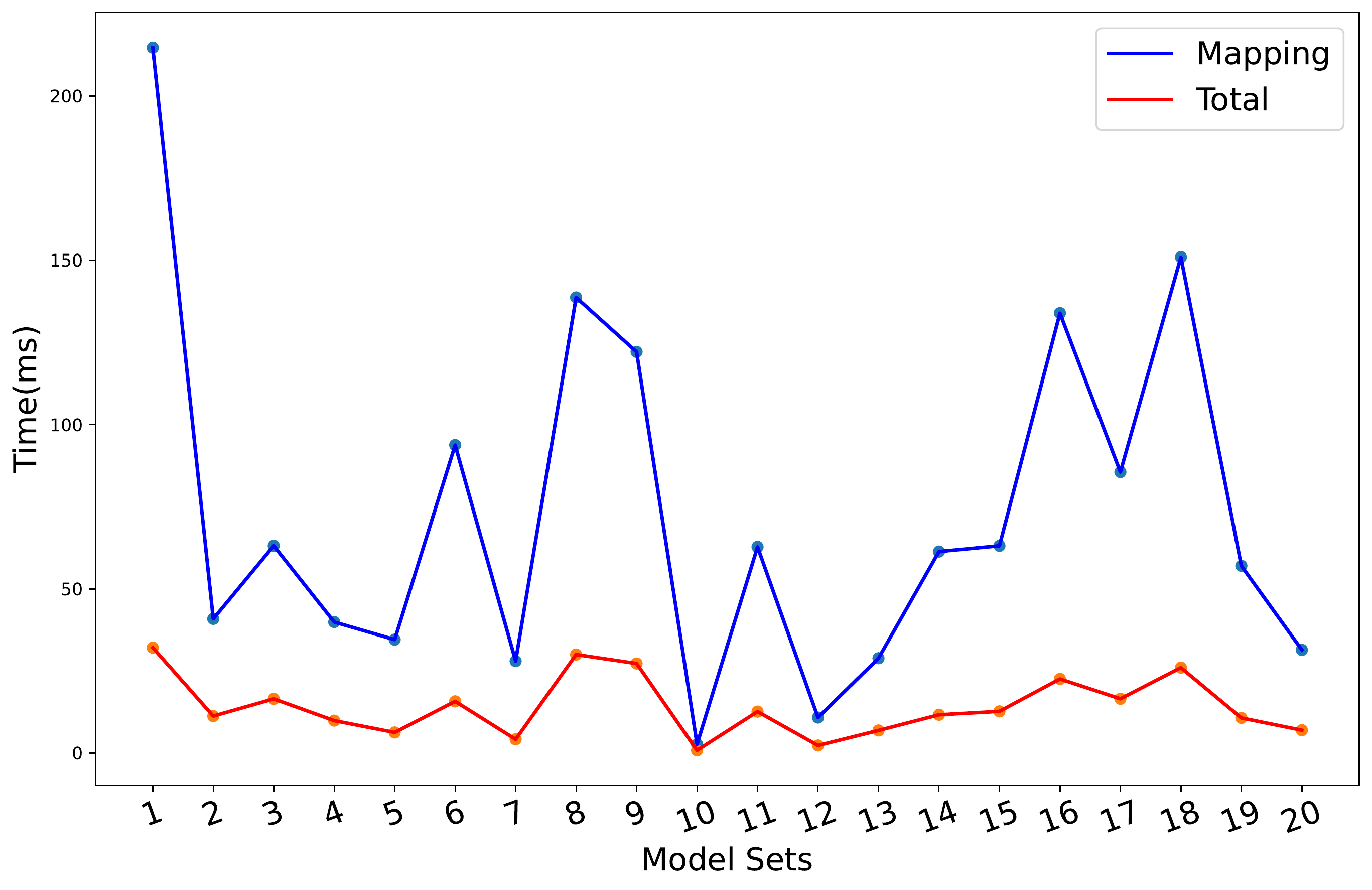}
    \caption{\centering Time-consuming comparison (BPEL)}
    \label{time_comparision_BPEL}
\end{figure}

Figures \ref{time_comparision_EPC} and \ref{time_comparision_BPEL} show the time consumption (Y axis) for co-evolving business process families in 20 model sets (X axis) in EPML and BPEL, respectively. The red and blue broken lines denote time consumption with our co-evolution method through a configurable process graph and the method for unmerged variants to individually propagate their changes, respectively. The figures show that in most cases, it takes much less time for our co-evolution method through a configurable process graph than for unmerged variants to directly propagate changes, indicating the high efficiency of our method.

As shown in Fig.\ref{time_comparision_BPEL}, the model set 10 exhibits no difference in time consumption between direct propagation and propagation after merging variants, which might be attributed to the small size of the model set 10. In practical applications, the direct change propagation time between variants increases with the increasing number of variants. In contrast, if the variant changes are firstly propagated to the configurable model, and then propagated to other variants, the time consumption will be greatly reduced, which might explain why the time consumption by our method is much less than that consumed by direct propagation. Our experiment results confirm that our method based on the co-evolution of process model families is more time-saving than the traditional methods, suggesting the high efficiency of our method.

\begin{figure}
    \centering
    \includegraphics[width=80mm]{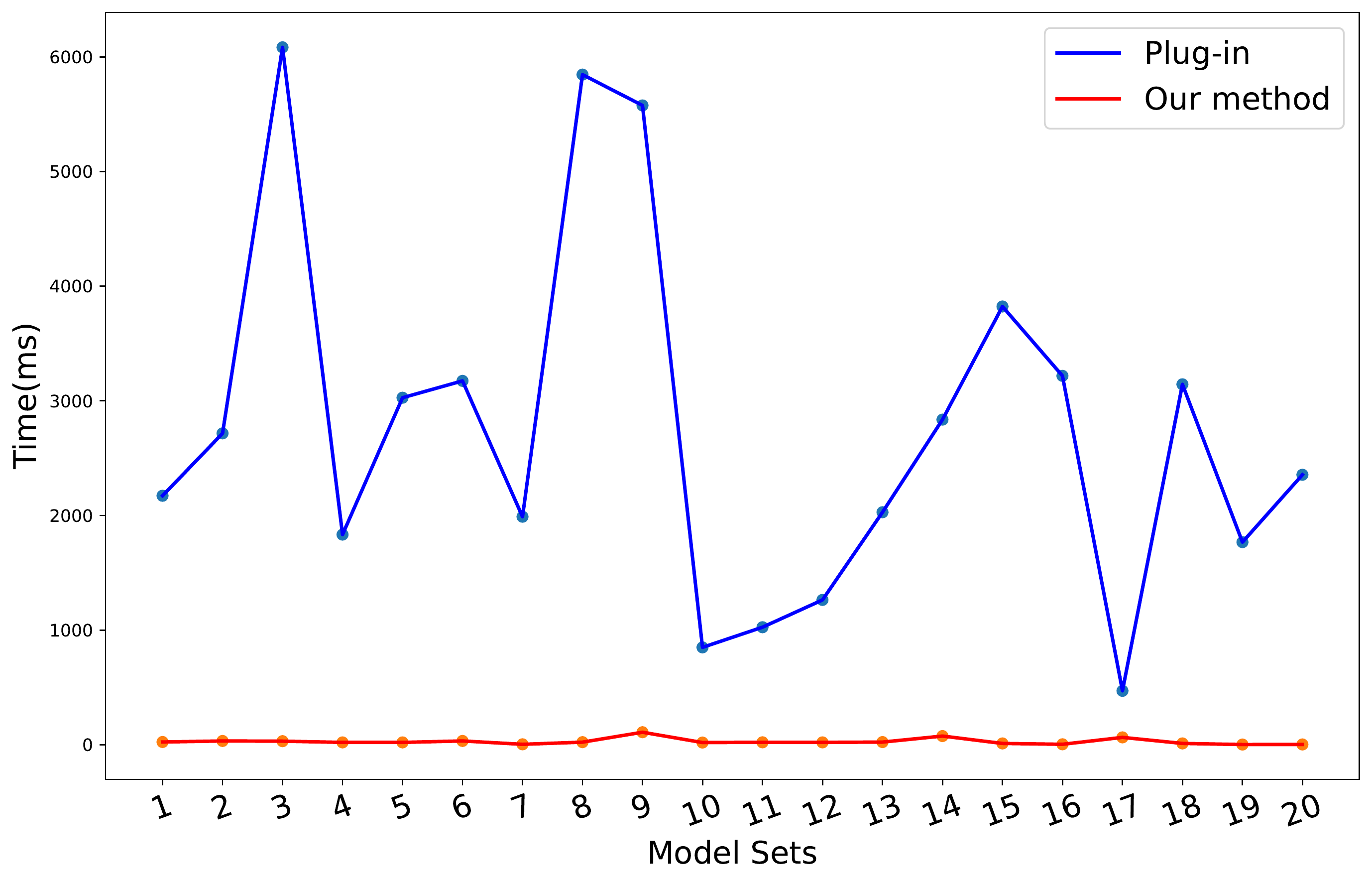}
    \caption{\centering Time-consuming comparison (Plug-in methods and operations)}
    \label{comparision_with_plug_in}
\end{figure}

Figure \ref{comparision_with_plug_in} summarizes the comparison between the \emph{plug-in} based co-evolution method \cite{r2} and our method using 20 EPC model sets of SAP dataset. The data showed that the correctness rate of the plug-in method (from 75\% to 100\%) is comparable to ours. However, the time consumption of the plug-in method is much higher than that of our method. The average time consumption of the plug-in method is more than 1 second, and the maximum time consumption is even 6 seconds, whereas the average time consumption of our method is tens of milliseconds. The reason for such a difference lies in that our co-evolution method can directly find the nodes and edges that need to be modified through a configurable process graph since the mappings between process variants and configurable process graph are \emph{pre-computed} during the process of merging. However, the \emph{variation} points need to be calculated with logical inferences in the plug-in method, which is much more complicated than ours.

\section{Conclusion}\label{sec7}

This paper proposes a co-evolution approach between process variants by configurable business process models. First, we define a series of process change primitives that act on both process variants and configurable business process models. Then, we design a bidirectional change propagation algorithm between process variants and configurable process models. Next, we propose a system architecture for the co-evolution of process variants, implement this system, and integrate the change propagation modules into Apromore. Finally, we conducted experiments on 155 commercial BPEL instance models and 604 R/3 reference business process models  \cite{r9}. The results show that our method can achieve the co-evolution of process variants with a high correctness rate and low time consumption. However, the detection and resolution of change propagation-induced inconsistencies of process models remain to be further investigated.

\section*{Acknowledgements}\label{sec8}
We appreciate Professor Marcello La Rosa in the School of Computing and Information Systems, The University of Melbourne, for his advice on early version of this article. This research project was supported in part by the Innovation fund of Chinese Marine Defense Technology Innovation Center under Grant JJ-2021-722-04, and in part by the Inner Mongolia Key Scientific and Technological Project under Grant 2021SZD0099, and in part by the Fundamental Research Funds for the Chinese Central Universities under Grant 2662020XXQD01, 2662022JC004.

\bibliographystyle{unsrt}
\bibliography{sn-bibliography}

\begin{thebibliography}{10}

\bibitem{r18}
Siddarth Ganesan, Young Yoon, and Hans-Arno Jacobsen.
\newblock Ni{\~n}os take five: the management infrastructure for distributed
  event-driven workflows.
\newblock In {\em Proceedings of the 5th ACM international conference on
  Distributed event-based system}, pages 195--206, 2011.

\bibitem{r15}
Wei Song, Fangfei Chen, Hans-Arno Jacobsen, Xiaoxu Xia, Chunyang Ye, and
  Xiaoxing Ma.
\newblock Scientific workflow mining in clouds.
\newblock {\em IEEE Transactions on Parallel and Distributed Systems},
  28(10):2979--2992, 2017.

\bibitem{r16}
Yousif Al~Ridhawi and Ahmed Karmouch.
\newblock Decentralized plan-free semantic-based service composition in mobile
  networks.
\newblock {\em IEEE Transactions on Services Computing}, 8(1):17--31, 2014.

\bibitem{rosa2017business}
Marcello~La Rosa, Wil MP Van~Der Aalst, Marlon Dumas, and Fredrik~P Milani.
\newblock Business process variability modeling: A survey.
\newblock {\em ACM Computing Surveys (CSUR)}, 50(1):1--45, 2017.

\bibitem{r1}
Marcello La~Rosa, Marlon Dumas, Reina Uba, and Remco Dijkman.
\newblock Business process model merging: An approach to business process
  consolidation.
\newblock {\em ACM Transactions on Software Engineering and Methodology
  (TOSEM)}, 22(2):1--42, 2013.

\bibitem{r48}
Xiang Gao, Yurong Chen, Zizhe Ding, Meng Wang, Xiaonan Zhang, Zhiqiang Yan,
  Lijie Wen, Qinlong Guo, and Ran Chen.
\newblock Process model fragmentization, clustering and merging: an empirical
  study.
\newblock In {\em International Conference on Business Process Management},
  pages 405--416. Springer, 2013.

\bibitem{r25}
Shamila Mafazi, Georg Grossmann, Wolfgang Mayer, and Markus Stumptner.
\newblock On-the-fly change propagation for the co-evolution of business
  processes.
\newblock In {\em OTM Confederated International Conferences" On the Move to
  Meaningful Internet Systems"}, pages 75--93. Springer, 2013.

\bibitem{r2}
Zaiwen Feng, Dickson~KW Chiu, Rong Peng, Ping Gong, Keqing He, and Yiwang
  Huang.
\newblock Facilitating cloud process family co-evolution by reusable process
  plug-in: An open-source prototype.
\newblock {\em IEEE Transactions on Services Computing}, 10(6):854--867, 2017.

\bibitem{r6}
Hoa~Khanh Dam and Aditya Ghose.
\newblock Mining version histories for change impact analysis in business
  process model repositories.
\newblock {\em Computers in Industry}, 67:72--85, 2015.

\bibitem{r7}
Matthias Weidlich, Jan Mendling, and Mathias Weske.
\newblock Propagating changes between aligned process models.
\newblock {\em Journal of Systems and Software}, 85(8):1885--1898, 2012.

\bibitem{r34}
Raminvas Laddad.
\newblock {\em Aspectj in action: enterprise AOP with spring applications}.
\newblock Simon and Schuster, 2009.

\bibitem{r4}
Michael Rosemann and Wil~MP Van~der Aalst.
\newblock A configurable reference modelling language.
\newblock {\em Information systems}, 32(1):1--23, 2007.

\bibitem{r9}
T~Curren and Gerhard Keller.
\newblock Sap r/3 business blueprint.
\newblock {\em Understanding the business process reference model. Upper Saddle
  River, NJ}, 1998.

\bibitem{r5}
Marcello La~Rosa, Hajo~A Reijers, Wil~MP Van Der~Aalst, Remco~M Dijkman, Jan
  Mendling, Marlon Dumas, and Luciano Garc{\'\i}a-Ba{\~n}uelos.
\newblock Apromore: An advanced process model repository.
\newblock {\em Expert Systems with Applications}, 38(6):7029--7040, 2011.

\bibitem{r3}
Georg Grossmann, Shamila Mafazi, Wolfgang Mayer, Michael Schrefl, and Markus
  Stumptner.
\newblock Change propagation and conflict resolution for the co-evolution of
  business processes.
\newblock {\em International Journal of Cooperative Information Systems},
  24(01):1540002, 2015.

\bibitem{weber2011}
Barbara Weber, Manfred Reichert, Jan Mendling, and Hajo~A Reijers.
\newblock Refactoring large process model repositories.
\newblock {\em Computers in industry}, 62(5):467--486, 2011.

\bibitem{r11}
Wei Song and Hans-Arno Jacobsen.
\newblock Static and dynamic process change.
\newblock {\em IEEE Transactions on Services Computing}, 11(1):215--231, 2016.

\bibitem{r47}
Pascal Poizat, Gwen Sala{\"u}n, and Ajay Krishna.
\newblock Checking business process evolution.
\newblock In {\em International Workshop on Formal Aspects of Component
  Software}, pages 36--53. Springer, 2016.

\bibitem{r40}
Damian Arellanes and Kung-Kiu Lau.
\newblock Workflow variability for autonomic iot systems.
\newblock In {\em 2019 IEEE International Conference on Autonomic Computing
  (ICAC)}, pages 24--30. IEEE, 2019.

\bibitem{r41}
Damian Arellanes and Kung-Kiu Lau.
\newblock D-xman: a platform for total compositionality in service-oriented
  architectures.
\newblock In {\em 2017 IEEE 7th International Symposium on Cloud and Service
  Computing (SC2)}, pages 283--286. IEEE, 2017.

\bibitem{r43}
Riccardo Cognini, Flavio Corradini, Stefania Gnesi, Andrea Polini, and Barbara
  Re.
\newblock Business process flexibility-a systematic literature review with a
  software systems perspective.
\newblock {\em Information Systems Frontiers}, 20(2):343--371, 2018.

\bibitem{r42}
Diana Kalibatiene and Olegas Vasilecas.
\newblock A survey on modeling dynamic business processes.
\newblock {\em PeerJ Computer Science}, 7:e609, 2021.

\bibitem{r36}
Andrea Delgado, Daniel Calegari, F{\'e}lix Garc{\'\i}a, and Barbara Weber.
\newblock Model-driven management of bpmn-based business process families.
\newblock {\em Software and Systems Modeling}, pages 1--37, 2022.

\bibitem{r39}
Daniel Calegari, Andrea Delgado, and Leonel Pe{\~n}a.
\newblock Automated generation of variants in business process families based
  on the common variability language (cvl).
\newblock In {\em 2019 XLV Latin American Computing Conference (CLEI)}, pages
  1--10. IEEE, 2019.

\bibitem{r38}
Daniel Calegari, Andrea Delgado, and Leonel Pe{\~n}a.
\newblock Model-driven support for business process families with the common
  variability language (cvl).
\newblock {\em CLEI Electron. J.}, 23(1):1--24, 2020.

\bibitem{r44}
Ramo {\v{S}}endelj and Ivana Ognjanovi{\'c}.
\newblock Multi-criteria decision making for optimal configuration of business
  process model families.
\newblock {\em Information technology and control}, 47(3):532--550, 2018.

\bibitem{r46}
Bedilia Estrada-Torres, Adela Del-R{\'\i}o-Ortega, Manuel Resinas, and Antonio
  Ruiz-Cort{\'e}s.
\newblock Modeling variability in the performance perspective of business
  processes.
\newblock {\em IEEE Access}, 9:111683--111703, 2021.

\bibitem{thomas2008}
Oliver Thomas.
\newblock Design and implementation of a version management system for
  reference modeling.
\newblock {\em J. Softw.}, 3(1):49--62, 2008.

\bibitem{r22}
Christian Gerth, Jochen~M K{\"u}ster, Markus Luckey, and Gregor Engels.
\newblock Detection and resolution of conflicting change operations in version
  management of process models.
\newblock {\em Software \& Systems Modeling}, 12(3):517--535, 2013.

\bibitem{r23}
Petra Brosch, Gerti Kappel, Martina Seidl, Konrad Wieland, Manuel Wimmer, Horst
  Kargl, and Philip Langer.
\newblock Adaptable model versioning in action.
\newblock {\em Modellierung}, 2010.

\bibitem{r20}
PN~Stuart~J Russell.
\newblock Artificial intelligence-a modern approach (3. internat. ed.), 2010.

\bibitem{r8}
Wei Song, Fangfei Chen, Hans-Arno Jacobsen, and Chengzhen Zhang.
\newblock Identifying a minimum sequence of high-level changes between
  workflows.
\newblock {\em IEEE Transactions on Services Computing}, 2021.

\bibitem{r17}
Chen Li, Manfred Reichert, and Andreas Wombacher.
\newblock On measuring process model similarity based on high-level change
  operations.
\newblock In {\em International Conference on Conceptual Modeling}, pages
  248--264. Springer, 2008.

\bibitem{r21}
Jochen~M K{\"u}ster, Christian Gerth, Alexander F{\"o}rster, and Gregor Engels.
\newblock Detecting and resolving process model differences in the absence of a
  change log.
\newblock In {\em International Conference on Business Process Management},
  pages 244--260. Springer, 2008.

\bibitem{r31}
Alexander Dreiling, Michael Rosemann, Wil M. P. Van~Der Aalst, Wasim Sadiq, and
  Sana Khan.
\newblock Model-driven process configuration of enterprise systems.
\newblock {\em DBLP}, 2005.

\bibitem{r32}
Alexander Dreiling, Michael Rosemann, Wil Van Der~Aalst, Lutz Heuser, and
  Karsten Schulz.
\newblock Model-based software configuration: patterns and languages.
\newblock {\em European Journal of Information Systems}, 15(6):583--600, 2006.

\bibitem{r33}
Marcello La~Rosa, Marlon Dumas, Arthur~HM Ter~Hofstede, and Jan Mendling.
\newblock Configurable multi-perspective business process models.
\newblock {\em Information Systems}, 36(2):313--340, 2011.

\bibitem{hallerbach2010capturing}
Alena Hallerbach, Thomas Bauer, and Manfred Reichert.
\newblock Capturing variability in business process models: the provop
  approach.
\newblock {\em Journal of Software Maintenance and Evolution: Research and
  Practice}, 22(6-7):519--546, 2010.

\bibitem{kumar2012design}
Akhil Kumar and Wen Yao.
\newblock Design and management of flexible process variants using templates
  and rules.
\newblock {\em Computers in Industry}, 63(2):112--130, 2012.

\bibitem{r19}
G.~Di Battista and R.~Tamassia.
\newblock On-line maintenance of triconnected components with spqr-trees.
\newblock {\em Algorithmica}, 4:302--318, 1996.

\bibitem{dijkman2011}
Remco Dijkman, Marlon Dumas, Boudewijn Van~Dongen, Reina K{\"a}{\"a}rik, and
  Jan Mendling.
\newblock Similarity of business process models: Metrics and evaluation.
\newblock {\em Information Systems}, 36(2):498--516, 2011.

\bibitem{bouguettaya2015}
Athman Bouguettaya, Qi~Yu, Xumin Liu, Xiangmin Zhou, and Andy Song.
\newblock Efficient agglomerative hierarchical clustering.
\newblock {\em Expert Systems with Applications}, 42(5):2785--2797, 2015.

\bibitem{r14}
T~Allweyer.
\newblock Bpmn 2.0, books on demand gmbh, norderstedt.
\newblock 2010.

\end{thebibliography}

\clearpage

\appendix

\section{Notations}
\bigskip
\bigskip
\bigskip
\bigskip

\resizebox{150mm}{80mm}{
\begin{tabular}{|c|l|}
\hline
{\bf Notation} & {\bf Meaning} \\\hline
$CG$ & Configurable graph obtained by merging $G_1$ and $G_2$. \\\hline
$G_1$, $G_2$ & Graphs to be merged. \\\hline
$v_p/v_s$ & The start/end node of an edge.\\
\hline
$\alpha_ G$ & The variant annotation of $G$.\\
\hline
$\alpha_ {CG}(v_p,v_s)$ & Annotation of the edge between $v_p$ and $v_s$ on model $CG$.\\\hline
$\alpha_ {(v_p,v_s)}$ & Annotation of the edge between $v_p$ and $v_s$.\\\hline
$\{p\}$ & Set of nodes in path $p$. \\\hline
$x \hookrightarrow y$ & Path from node $x$ to node $y$. \\\hline
$x \stackrel{c}{\hookrightarrow} y$ & Connector chain, i.e.\ path of connectors from node $x$ to node $y$. \\\hline
$p(v)/s(v)$ & Parent/Child node of v. \\\hline
$\gamma_G(x)$ & \tabincell{l}{Returns the annotation of node $x$ in graph $G$, i.e. the set of pairs $(pid,l)$\\ where $pid$ is a process graph identifier and $l$ is the label of $x$ in graph $pid$.} \\\hline
$\eta_G(x)$ & Returns true if connector $x$ is configurable, false otherwise. \\\hline
$\lambda_G(x)$ & Returns the label of node $x$ in graph $G$. \\\hline
$\tau$(x) &  \tabincell{c}{The type of node x. There are three types: (e)vent, (f)unction, and (c)onnector.} \\\hline
$m=(c, xor)$ & Create a connector m of type xor. \\\hline
$M_{G,CG}^N(v)$ & Returns the node in $CG$ that is matched with the node $v$ in $G$ \\\hline
$M_{G,CG}^E(e)$ & Returns the edge in $CG$ that is matched with the edge $e$ in $G$ \\\hline
\emph{mcrs} & Set of maximum common regions between $G_1$ and $G_2$ \\\hline
$N_{CG}$ & Node set for model $CG$. \\\hline
$E_{CG}$ & Edge set for model $CG$. \\\hline
\end{tabular}}

\clearpage

\end{document}